\documentclass{svmult}

\usepackage{makeidx}         
\usepackage{graphicx,cite}        
\usepackage{multicol}        
\usepackage[bottom]{footmisc}

\usepackage{amssymb,amsmath,version,color}
\usepackage{float}
\newcommand{\ddr}{\textrm{d}}
\newcommand{\dd}{\textrm{d}}
\newcommand{\ee}{\text{e}}
\newcommand{\bv}{{\bf v}}
\newcommand{\vvec}{{\bf v}}

\newcommand{\rvec}{{\bf r}}
\newcommand{\bc}{{\bf c}}

\newcommand{\bF}{{\bf F}}

\newcommand{\mut}{{\tilde \mu}}
\newcommand{\lambt}{{\tilde \lambda}}
\newcommand{\lambdat}{{\tilde \lambda}}
\newcommand{\vt}{{\tilde v}_0}
\newcommand{\ft}{{\tilde f}}
\newcommand{\pit}{{\tilde \pi}}
\newcommand{\wt}{{\tilde w}}
\newcommand{\cWt}{{\tilde {\cal W}}}
\newcommand{\cW}{{\cal W}}

\newcommand{\cO}{{\cal O}}

\newcommand{\bsig}{\hat{\boldsymbol \sigma}}
\newcommand{\svec}{\hat{\boldsymbol \sigma}}

\newcommand{\Mvariable}{{}}
\newcommand{\p}{\partial}
\newcommand{\bo}{\hat{\boldsymbol \omega}}
\newcommand{\bs}{\hat{\boldsymbol \sigma}}
\newcommand{\im}{{{\textrm i}}}
\graphicspath{{figures/}}

\makeindex             

\begin{document}

\title*{Fluctuations in granular gases}
\author{A. Barrat\inst{1}\and A. Puglisi\inst{2}\and
 E. Trizac\inst{3}\and P. Visco\inst{4}\and F. van Wijland\inst{5}}
\institute{Laboratoire de Physique Th\'eorique (CNRS UMR 8627),
  B\^atiment 210, Universit\'e Paris-Sud, 91405 Orsay cedex, France
  \texttt{Alain.Barrat@th.u-spud.fr} \and Dipartimento di Fisica,
  Universit\`a La Sapienza, p.le A. Moro 2, 00185 Roma, Italia
  \texttt{andrea.puglisi@roma1.infn.it} \and Laboratoire de Physique
  Th\'eorique et Mod\`eles Statistiques (CNRS UMR 8626), B\^atiment
  100, Universit\'e Paris-Sud, 91405 Orsay cedex, France
  \texttt{trizac@ipno.in2p3.fr} \and Laboratoire de Physique
  Th\'eorique et Mod\`eles Statistiques (CNRS UMR 8626), B\^atiment
  100, Universit\'e Paris-Sud, 91405 Orsay cedex, France
  \texttt{Paolo.Visco@th.u-spud.fr} \and Laboratoire Mati\`ere et
  Syst\`emes Complexes (CNRS UMR 7057), Universit\'e Denis Diderot
  (Paris VII), 2 place Jussieu, 75251 Paris cedex 05, France
  \texttt{Frederic.Van-Wijland@th.u-psud.fr}}
%
%
\maketitle

\begin{abstract}
  A driven granular material, e.g. a vibrated box full of sand, is a
  stationary system which may be very far from equilibrium. The standard
  equilibrium statistical mechanics is therefore inadequate to describe
  fluctuations in such a system. Here we present numerical and analytical
  results concerning energy and injected power fluctuations. In the first part
  we explain how the study of the probability density function (pdf) of the
  fluctuations of total energy is related to the characterization of velocity
  correlations. Two different regimes are addressed: the gas driven at the
  boundaries and the homogeneously driven gas. In a granular gas, due to
  non-Gaussianity of the velocity pdf or lack of homogeneity in hydrodynamics
  profiles, even in the absence of velocity correlations, the fluctuations of
  total energy are non-trivial and may lead to erroneous conclusions about the
  role of correlations.  In the second part of the chapter we take into
  consideration the fluctuations of injected power in driven granular gas
  models. Recently, real and numerical experiments have been interpreted as
  evidence that the
  fluctuations of power injection seem to satisfy the Gallavotti-Cohen Fluctuation
  Relation. We will discuss an alternative interpretation of such results
  which invalidates the Gallavotti-Cohen symmetry. Moreover, starting from
  the Liouville equation and using techniques from large deviation theory, the general
  validity of a Fluctuation Relation for power injection in driven granular
  gases is questioned. Finally a functional is defined using the
  Lebowitz-Spohn approach for Markov processes applied to the linear inelastic
  Boltzmann equation relevant to describe the motion of a tracer particle. 
  Such a functional results to be different from injected
  power and to satisfy a Fluctuation Relation.
\end{abstract}
\section{Introduction}
\label{intro}

In equilibrium thermodynamics one characterizes the stable phases of a system 
using a limited set of macroscopic state variables, therefore bypassing much of 
the microscopic details of the systems under study. It is only very recently 
that the same strategy has been applied to systems in Non-Equilibrium 
Steady-States (NESS). And the need for such an approach is all the more pregnant 
for the study of NESS that no general formalism parallel to the standard 
equilibrium Gibbs-Boltzmann ensemble theory exists. This field started with 
experiments carried out on turbulent flows or convection cells, and much more 
recently on granular systems. Global observables, namely spatially integrated 
over the whole system, and their distribution, may indeed coarse-grain the 
irrelevant microscopic details specific to the system at hand, while allowing 
for comparisons between different systems. They are expected to be more robust 
and more exportable tools for analysis than local probes, like {\it e.g.} 
structure factors. This has led to the
observation of intriguing similarities between 
turbulent flows and granular systems~\cite{bramwell98,brey05,bertin05}. However, one 
must take into account the key ingredient making a NESS way different from its 
equilibrium counterpart: steady flows of energy, matter, or else, run across the 
system. The existence of currents characterizes a NESS, and makes it different 
from an equilibrium state in that detailed balance (time reversibility) no 
longer holds. Given that the time direction plays a central r\^ole, one is led to 
the idea that time integration may also be useful in smoothening out various 
details of the microscopic dynamics. This has motivated several authors to 
consider the distribution of time integrated and spatially averaged quantities 
characterizing the NESS as such, like that of the injected power in a turbulent 
flow or in a granular gas.\\ 

We briefly turn to a reminder of phenomenological thermodynamics of 
nonequilibrium systems, as presented in \cite{degroot}. There, for systems only 
slightly away from equilibrium, the concept of entropy can be extended in a 
consistent fashion, and its time evolution goes according to
\begin{equation}\label{evolentropy}
\frac{dS}{dt}=\int_V\sigma_{\text{irr}}-\int_V\nabla\cdot{\bf J}_S.
\end{equation}
The intrinsic entropy production rate $\sigma_{\text{irr}}$ is positive 
definite, and cancels under the condition that the system reaches equilibrium. 
The other piece in the rhs of (\ref{evolentropy}), which features an entropy 
current ${\bf J}_S$, conveys the existence of external sources, often located at 
the system's boundaries, driving the system out of equilibrium. The entropy 
current is not but a linear combination of the various currents flowing through 
the system, with the conjugate affinities (like a temperature or a chemical 
potential gradient) as the proportionality factors. The entropy current -- when 
it can univoquely be defined -- therefore stands as a relevant measure of how far 
the system is from equilibrium. For that reason, various studies, starting from 
the pioneering work of Evans, Cohen and Morriss~\cite{evans93}, in a study of a 
thermostatted fluid under shear, have been focused on the appropriately 
generalized expression of the latter entropy current. In Evans, Cohen and 
Morriss' case it is simply proportional to the power provided by the thermostat 
to compensate for viscous dissipation. They went on to determine the 
distribution function of $Q_S(t)$, the time integrated entropy flow (or 
equivalently the energy provided by the thermostat), denoted by $P(Q_S,t)$. In 
doing so they empirically noticed a remarkable property of the pdf of $Q_S$, 
namely
\begin{equation}
\label{GCsym}
\lim_{t\to\infty}\frac{1}{t}\log\frac{ P(Q_S,t)}{P(-Q_S,t)}=q_S,
\end{equation}
where $q_S=Q_S/t$, which is a time-intensive quantity, is the time average of 
$J_S$ over $[0,t]$. This symmetry property of the pdf of $Q_S$ was soon to be 
formalized into a theorem for thermostatted systems by Gallavotti and 
Cohen~\cite{gallavotti95}, and has since triggered a flurry of studies. The mathematical object defined by $\pi(q_S)=\lim_{t\to\infty}\frac{\log 
P(q_S \;t,t)}{t}$ is seen to be extending the concept of intensive free 
energy to a nonequilibrium setting, and will occupy much of our numerical and 
analytical efforts.\\

In the realm of nonequilibrium systems, granular gases play a central r\^ole as 
systems exhibiting a strongly irreversible microscopic dynamics due to inelastic 
collisions, and for these no viable definition of entropy, let alone entropy 
flow, is available. This has led various authors~\cite{aumaitre01,feitosa04} to 
conjecture that, by analogy to thermostatted systems, the power injected into 
the system to maintain it in a steady-state, could satisfy a 
symmetry property like the one uncovered by~\cite{evans93}, and its ensuing 
consequences in terms of generalized fluctuation--dissipation theorems. 
Fortunately, a well-controlled kinetic theory-based statistical mechanics exists 
 for dilute gaseous systems, and we shall build upon it to investigate the 
questions raised above.\\

The outline of the present review is as follows. We begin in Sec. 2 with a brief 
introduction to granular gases and the basics of their statistical mechanics. In 
Sec. 3 we analyze the distribution of the total kinetic energy of the gas, as a 
first choice for a global observable. In Sec. 4 and 5 we address numerically, 
analytically and also experimentally, the issue of interpreting the power 
injected into a gas in terms of entropy flow, the negative outcome of which 
leads us to Sec. 6. There we construct a one particle observable  exhibiting the 
properties expected from an entropy flow, and quite notably its distribution 
function displays the symmetry property (\ref{GCsym}).

\section{A brief introduction to granular gases}
\label{gases}

A granular gas is an assembly of macroscopic particles kept in a
gaseous steady state by a constant excitation~\cite{barrat05}(the
typical example can be illustrated thinking of many beads in a
strongly vibrated box). The simplest way to characterize such systems
is to consider $N$ identical smooth hard spheres, losing a part of
their kinetic energy after each collision.  The total momentum is
conserved in collisions, and only the normal component of the velocity
is affected. Thus, the collision law for a couple of particles $(1,2)$
reads:
\begin{equation}
\label{coll}
\begin{cases}
  {\bf v}_1^*={\bf v}_1-{1 \over 2} (1+\alpha)({\bf v}_{12} \cdot {\hat
    {\boldsymbol \sigma}})
  {\hat {\boldsymbol \sigma}} \\
  {\bf v}_2^*={\bf v}_2+{1 \over 2} (1+\alpha)({\bf v}_{12} \cdot {\hat
    {\boldsymbol \sigma}})
  {\hat {\boldsymbol \sigma}}  \,\, ,\\
 \end{cases}
 \end{equation}
 where $\svec$ is a unitary vector along the center of the colliding
 particles at contact. Here $\alpha$ is a constant, called the
 coefficient of normal restitution ($0\le\alpha\le1$, and when
 $\alpha=1$ collisions are purely elastic). Without some energy
 injection mechanism the total energy of the gas will decrease in
 time, until all the particles are at rest (cooling state). However,
 when some energy input is provided, the system can reach a
 nonequilibrium stationary state. Energy injection may be supplied in
 several ways, which can be divided in two main categories: injection
 from the boundaries and homogeneous driving. In the former category
 energy is supplied by a boundary condition, the system hence develops
 spatial gradients and it is not homogeneous. The latter
category refers to systems where energy injection is achieved by 
a homogeneous and isotropic force acting on each particle.

\subsection{Boundary driven gases}
In this section we will give a short introduction to the methods used
to describe the behavior of a granular gas in which energy is injected
by a boundary condition (typically a vibrating wall).  This kind of
system has been widely studied in the literature
\cite{grossman97,mcnamara97,mcnamara98b,kumaran98,brey00c,barrat02b},
and one of its main characteristics is that the density and the
temperature are not homogeneous over the system: there is a heat flux,
which does not verify Fourier law.  This feature is well described by
kinetic theory and in good agreement with the hydrodynamic
approximation, which allows an analytical calculation of the density
and temperature profiles. In the dilute limit, such a system is well
described by the Boltzmann equation:
\begin{equation}
\label{boltzpos}
\partial_t f(\rvec,\vvec_1, t)+ \vvec_1 \cdot {\boldsymbol \nabla} 
f(\rvec,\vvec_1, t) = J[f|f]\,\,.
\end{equation}   
Here $J[f|f]$ is the collision integral, which takes into account the
inelasticity of the particles:
\begin{multline}
\label{collint}
J[f|f]=\sigma^{d-1} \int \ddr \vvec_2 \int' \ddr \svec (\vvec_{12} \cdot
\svec) \left(\frac{f(\vvec_1^{**},t)f(\vvec_2^{**},t)}{\alpha^2}
- f(\vvec_1,t) f(\vvec_2,t) \right),
\end{multline}
where the notation $\bv_{12}$ denotes the relative velocity between
particles 1 and 2, the two stars superscript (i.e. $\bv^{**}$) denote
the precollisional velocity of a particle having velocity $\bv$, and
the primed integral is a short-hand notation meaning that the
integration is performed on all angles satisfying
$\mathbf{v}_{12}\cdot\bsig>0$. The hydrodynamic fields
are defined as the velocity moments:
\begin{equation}
n(\rvec,t)=\int \ddr \vvec f(\rvec,\vvec,t)\,,
\end{equation}
\begin{equation}
n(\rvec,t) {\bf u}(\rvec,t) = \int \ddr \vvec \, \vvec f(\rvec,\vvec,t) \,,
\end{equation}
\begin{equation}
{d \over 2} n(\rvec,t) T(\rvec,t)= \int \ddr \vvec {m \over 2} 
(\vvec - {\bf u})^2 f(\rvec,\vvec,t)\,,
\end{equation}
and the hydrodynamic balance equations for those quantities are derived taking
the velocity moments in the equation (\ref{boltzpos}). Their expression is:
\begin{equation}
\partial _{t}n+\nabla \cdot \left( n\mathbf{u}\right) =0,  \label{2.5a}
\end{equation}
\begin{equation}
\left( \partial _{t}+\mathbf{u}\cdot \nabla \right)
u_{i}+(mn)^{-1}{\nabla}_{j}P_{ij}=0,  \label{2.5b}
\end{equation}
\begin{equation}
\left( \partial _{t}+\mathbf{u}\cdot \nabla +\zeta \right)
T+\frac{2}{3n} \left( P_{ij}{\nabla }_{j}u_{i}+\nabla \cdot
\mathbf{q}\right) =0, \label{2.5c}
\end{equation}
where the pressure tensor $P_{ij}$, heat flux $\mathbf{q}$, and the cooling
rate $\zeta$ are defined by:
\begin{equation}
P_{ij}(\mathbf{r},t)=\int d\mathbf{v}\,m (v_{i}
-u_{i})(v_{j}-u_{j})f(\mathbf{r},\mathbf{v},t)\,\,, \label{2.5d}
\end{equation}
\begin{equation}
\mathbf{q}(\mathbf{r},t)=\int
d\mathbf{v}\frac{m}{2}(\mathbf{v}-\mathbf{u})^{2}
(\mathbf{v}-\mathbf{u}) f(\mathbf{r},\mathbf{v},t)\,\,, \label{2.5e}
\end{equation}
\begin{equation}
\zeta (\mathbf{r},t)=\frac{(1-\alpha^{2}) m\pi^{\frac{d-1}{2}} \sigma
  ^{d-1}}{4 d \Gamma\left(\frac{d+3}{2}\right) n(\mathbf{r},t)T(\mathbf{r},t)}
\int \,\dd\mathbf{v}_1\,\int \,\dd\mathbf{v}_{2}\,|v_{12}|^{3}\
f(\mathbf{r},\mathbf{v}_1,t)f(\mathbf{r},\mathbf{v}_{2},t).
\label{2.5}
\end{equation}
Explicit analytical expressions for the above quantities have been
obtained in the limit of small spatial gradients by Brey {\it et al.}
\cite{brey98,brey01d}. Moreover for systems in the steady state
without a macroscopic velocity flow the hydrodynamic equations
simplify, and therefore the temperature and density profiles can be
explicitly computed.

\subsection{Randomly driven gases} \label{randomlydriven}
We consider here a granular gas kept in a stationary state by an
external homogeneous thermostat, the so called ``Stochastic
thermostat'', which couples each particle to a white noise. Energy
injection is hence achieved by means of random forces acting
independently on each particle, and drives the gas into a
non-equilibrium steady state. The equation of motion governing the
dynamics of each particle is therefore:
\begin{equation}
m {\ddr \bv_i \over \ddr t} = \bF_i^{\text{coll}} + \bF_i^{\text{th}}, 
\end{equation}
where $\bF_i^{\text{coll}}$ is the force due to collisions and
$\bF_i^{\text{th}}$ is a Gaussian white noise (i.e. $\langle
F_{i\gamma}^{\text{th}}(t)F_{j\delta}^{\text{th}}(t')\rangle=2 \Gamma
\delta_{ij} \delta_{\gamma\delta}\delta(t-t')$, where the subscripts $i$ and
$j$ are used to refer to the particles, while $\gamma$ and $\delta$ denote the
Euclidean components of the random force).  This model is one of the most
studied in granular gas theory and reproduces many qualitative features of
real driven inelastic
gases~\cite{williams96b,peng98,puglisi98,noije99,henrique00b,moon01,pagonabarraga02,noije98c}.
After a few collisions per particle the system attains a non-equilibrium
stationary state. This state seems homogeneous. From the equations of motion it is possible to derive the
homogeneous Boltzmann equation governing the evolution of the one-particle
velocity distribution function~\cite{noije98c}:
\begin{equation}
\label{boltz}
\partial_t f(\bv_1, t) = J[f,f] +\Gamma \Delta_{\bv_1} f(\bv_1)\,\,,
\end{equation}   
where the Laplace operator $\Delta_{\bv}\equiv (\p / \p_{\bv})^2$ is a
diffusion term in velocity space characterizing the effect of the
random force, while $J[f,f]$ is the collision integral, which takes
into account the inelasticity of the collisions (cf. eq.
(\ref{collint})). The granular temperature of the system is defined
as usual as the mean kinetic energy per degree of freedom,
$T_g=\langle v^2 \rangle /d$. The stationary solution of
equation~(\ref{boltz}) has extensively been investigated in the last
years. Even if an exact solution is still missing, a general method is
to look for solutions in the form of a Gaussian distribution
multiplied by a series of Sonine polynomials~\cite{chapman60}:
\begin{equation}
\label{stationarypdf}
  f_{st}(\bv)=e^{-{v^2 \over 2 T_g }} 
  \left(1+ \sum_{p=1}^{\infty} a_p
    S_p\left({v^2 \over 2 T_g} \right) \right)\,\,.
\end{equation}
The expression of the first three Sonine polynomials is:
\begin{gather}\label{soninep}
  S_0(x)=1   \notag \\
  S_1(x)=-x + {1 \over 2} d  \\
  S_2(x)={1 \over 2} x^2 -{1 \over 2} (d+2)x + {1 \over 8} d(d+2) \,\,.\notag 
\end{gather}
Moreover the coefficients $a_p$ are found to be proportional to the averaged
polynomial of order $p$:
\begin{equation}
\label{coefficients}
a_p=A_p \left\langle S_p \left({v^2 \over 2 T_g} \right) \right\rangle\,\,,
\end{equation}
where $A_p$ is a constant and the angular brackets denote average 
with weight $f_{st}$. From this observation one directly obtains that the
first coefficient $a_1$ vanishes by definition of the temperature.  A first
approximation for the velocity pdf is therefore to truncate the expansion up
to the second order ($p=2$).  An approximated expression for the coefficient
$a_2$ has been found as a function of the restitution coefficient $\alpha$ and
the dimension $d$ \cite{noije98c, coppex03, montanero00b}. Its
expression is:
\begin{equation}
\label{a2}
a_2 (\alpha)={16 (1- \alpha)(1-2 \alpha^2) \over 73 +56 d 
-24 \alpha d -105 \alpha +30 (1-\alpha)\alpha^2}\,\,.
\end{equation}
It must be noted that the second Sonine approximation is only valid
for not too large velocities, since the tails of the pdf have been
shown ~\cite{noije98c} to be overpopulated with respect to the
Gaussian distribution. It is known~\cite{noije98c} that in high
energies $\log f(v) \sim -(v/v_c)^{3/2}$ with a threshold velocity
$v_c$ that diverges when the dimension $d$ goes to infinity. This
means that at high dimensions the distribution is almost a Gaussian,
since both the tails and the $a_2$ contributions tend to vanish. All
the above results have been confirmed by numerical simulations, in
particular through Molecular Dynamics (MD) and Direct simulation Monte
Carlo (DSMC) \cite{bird94} methods. Those two numerical methods,
although very different, show a surprisingly good agreement. This
points out to correctness of the molecular chaos assumption and thus
to the relevance of the DSMC method, which is particularly well
adapted to simulate the dynamics of a homogeneous dilute gas.

\section{Total energy fluctuations in vibrated and driven granular gases}
\label{efluct}

\subsection{The inhomogeneous boundary driven gas}
In this section we will study the energy fluctuations of a granular
gas in the case where the energy is injected into the system by a
vibrating wall.  Recently Auma\^{\i}tre {\it et al.}  \cite{aumaitre04}
investigated, by means of Molecular Dynamic (MD), the fluctuations of
the total energy of the system. In particular they looked at the
behavior of the first two moments of the energy pdf when the system
size is changed, at constant averaged density.  Because of the
inhomogeneities, the mean kinetic energy is no more proportional to
the number of particles, and thus it is not an extensive quantity, and
analogously the mean kinetic temperature is no more intensive.  This
has led to the definition of an effective (intensive) temperature and
an effective number of particles, which makes the energy extensive.
In the following we will show how a rough calculation (neglecting
correlations and small non-Gaussianity) using the hydrodynamic
prediction for the temperature profile~\cite{viscoepj}, can explain the phenomenology
observed in \cite{aumaitre04}. Within this description it is possible
to get an expression of the effective temperature and number of
particles as a function of the system parameters (i.e. number of
particles, restitution coefficient, and temperature of the vibrating
wall).
\subsubsection{Energy Probability Distribution Function}
In this part we will compute the energy pdf for a granular gas between
two (infinite) parallel walls. The distance between the two walls is
denoted by $H$, oriented along the $x$ axis. Here we assume that one of the walls (in
$x=0$) has small and random vibrations, acting as a thermostat that fixes to
$T_0$ the temperature at $x=0$. Our boundary conditions therefore are:
\begin{equation}
T(\ell=0)=T_0 \,\,,\,\,\,\,\,\,\,\,\,\,
\left.\frac{\partial T}{\partial \ell}\right|_{\ell=\ell_m}=\,\,
\left.\frac{\partial T}{\partial x}\right|_{x=H}=0\,\,,
\end{equation}
where the rescaled length $\ell$ will be defined below.
For the particular case of a steady state
without macroscopic velocity flow, is it possible to solve those
balance equations and get the temperature profile \cite{brey00c}:
\begin{equation}
\label{tempprofile}
T(\ell)=T_0 \left( {\cosh {\left( \sqrt{a(\alpha)} (\ell_m- \ell) \right)}
    \over \cosh{\left( \sqrt{a(\alpha)} \ell_m \right)}} \right)^2 \,\,,
\end{equation}
where $a(\alpha)$ is a function of the restitution coefficient (its complete
expression is given in ref. \cite{brey04a}). The variable $\ell$ is
proportional to the integrated density of the system on the $x$ axis. Its
definition is given by the following relation involving the local
mean-free-path $\lambda (x)$:
\begin{equation}
\label{lscale}
\textrm{d}\ell={\textrm{d} x \over \lambda (x)} \,\,, \lambda(x)= \left[ {\sqrt{2}
  \pi^{d-1 \over 2} \over \Gamma[(d+1)/2]} \sigma^{d-1} n(x) \right]^{-1}\,\,.
\end{equation}
In the following we will suppose the velocity distribution to be a Maxwellian
(a small non-Gaussian behavior exists, but it is not relevant for this
calculation) with a local temperature (variance) given by (\ref{tempprofile}):
\begin{equation}
f(\bv,\ell)={e^{-{v^2 \over 2 T(\ell)}} \over {(2 \pi T(\ell))^{d/2}}} \,\,.
\end{equation}
The distribution for the energy of one particle ($e=v^2/2$) is hence:
\begin{equation}
p(e,T(\ell))=f_{{1 \over T(\ell)}, {d \over 2}}(e) \,\,\,,
\end{equation}
where $f_{\alpha, \nu} (x)$ is the gamma distribution \cite{feller}:
\begin{equation}
\label{gamma}
f_{\alpha, \nu}(x)={\alpha^{\nu} \over \Gamma(\nu)} 
x^{\nu-1} e^{- \alpha x}\,. 
\end{equation}
Our interest goes to the macroscopic fluctuations integrated over all the
system. Thus, the macroscopic variable of interest is the granular temperature
$T_g$, defined here as the average of the local temperature over the $x$
profile:
\begin{equation}
\label{tgran}
  T_g={1 \over N}\int_V  n(\rvec) T(\rvec) \,\,\dd \rvec = {1 \over \ell_m} \int_{0}^{\ell_m} T(\ell) \,\,\textrm{d} \ell \,\,.
\end{equation}
with 
\begin{equation}
\ell_m = N_x \frac{
  \sqrt{2}\pi^{\frac{d-1}{2}}\sigma^{d-1}}{\Gamma[(d+1)/2]}, \;\;\; N_x=\frac{N}{V_{d-1}},
\end{equation}
where $V_{d-1}$ is the area of the surface of dimension $d-1$ orthogonal to
the $x$-direction, i.e. $H \times V_{d-1} = V$. When $d=2$ one has $V_{d-1}
\equiv L$ where $L$ is the width of the system. $N_x$ is the number of
particles per unit of section perpendicular to the $x$ axis.
To get an expression of the energy pdf over the whole system, it is useful to
divide the box in $\ell_m / \Delta \ell$ boxes of equal height (in the $\ell$
scale) $\Delta \ell$. It is helpful to use the length scale $\ell$ because the
number of particles $N_{\ell}$ in each box of size $L \times \Delta \ell$ is a
constant. Moreover, in each box $i$ we will suppose the temperature a constant
$T_i \equiv T(i \,\, \Delta \ell)$, defined expanding the granular temperature
in a Riemann sum:
\begin{equation}
T_g= \lim_{\Delta \ell \to 0} \sum_{i=0}^{\ell_m / \Delta \ell} T_i
\,\,\Delta \ell 
\end{equation}
The calculation of the pdf of the box energy $\epsilon_i$, i.e. a sum of the
energies of the $N_{\ell}$ particles in a box $i$, is hence straightforward
when the velocities of the particles are supposed to be uncorrelated:
\begin{equation}
q_i(y)\equiv \textrm{prob}(\epsilon_i=y)=f_{{1 \over T_i}, {dN_{\ell} \over 2}}(y) \,\,\,,
\end{equation}
The characteristic function of $q_i(y)$ is 
\begin{equation}
\tilde{q}_i(k)=\frac{1}{(1-ikT_i)^{\frac{dN_{\ell}}{2}}}
\end{equation}
Thus, the characteristic function for the  kinetic energy of the whole system $E= \sum
\epsilon_i$ can be obtained as the product of the characteristic functions
${\tilde q}_i(k)$:
\begin{equation}
\label{prod}
{\widetilde P}(k)= \prod_{j=0}^{\ell_m / \Delta \ell} {\tilde q}_j(k)=
\prod_{j=0}^{\ell_m / \Delta \ell} { 1 \over  (1 - \im k
  T_j)^{dN_{\ell} \over 2}} \,\,\,.
\end{equation}
Since the number of particle per box $N_{\ell}$ is a known fraction of the
total number of particles ($N_{\ell} = N \Delta \ell / \ell_m$), one can
rewrite the expression (\ref{prod}) as a Riemann sum. In the limit $\Delta
\ell \to 0$ this yields the total kinetic energy characteristic function:
\begin{equation}
\label{charfunen}
{\widetilde P}(k)= \exp  \left(\, -{d N \over 2 \ell_m } \int_0^{\ell_m} \log
\left(1-\im k T(\ell)\right) \,\,  \textrm{d} \ell \right)\,\,.  
\end{equation}
Note that this result is valid for any temperature profile $T(\ell)$ and hence
it can be applied also to other situations with different boundary conditions
or different hydrodynamic equations.

\subsubsection{Comparison with simulations}

\begin{figure}
  \centering
  \includegraphics[clip=true,width=0.45 \textwidth]{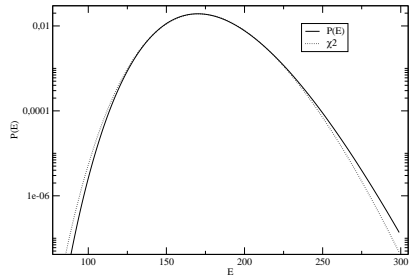}
  \caption{\label{figenpdf} Energy pdf (solid line) and a gamma distribution
    with same mean and same variance (dotted line) for a restitution
    coefficient $\alpha=0.9$ , $N=100$ particles in two dimensions in a box of
    density $\rho = 0.04$, height $H=50$, and with a wall temperature $T_0=5$.}
\end{figure}

Auma\^{\i}tre {\it et al.}  \cite{aumaitre01,aumaitre04} showed by Molecular
dynamic simulations, that the pdf of the total energy is well fitted by a
$\chi^2$ law $\Pi(E) = f_{{1 \over T_E}, {N_f \over 2}} (E)$
with a number of degrees of freedom $N_f$ different from $d N$, and a
temperature $T_E$ different from the granular temperature $T_g$. The two
parameters $N_f$ and $T_E$ are functions of the first two cumulants of the pdf:
\begin{equation}
N_f= 2 \, { \langle E \rangle_c^2  \over {\langle E^2 \rangle_c}}\,, 
\;\;\;\;\;\;\;\;\;\;\;\;\;\;\;\;\;\;\;\;
T_E= {\langle E^2 \rangle_c \over \langle E \rangle_c}\,\,.
\end{equation}
The notation $\langle X \rangle_c$ denotes the cumulant of the
variable $X$.  Here we want to compare result~\eqref{charfunen} with these
numerical results. Since we are not able to analytically calculate the inverse
Fourier Transform of~\eqref{charfunen} using~\eqref{tempprofile} as a
temperature profile, we used a numerical computation to obtain it in an
approximate form. Moreover, an expression of the cumulants of the total
kinetic energy can be obtained from the characteristic function
(\ref{charfunen}):
\begin{equation} 
\label{energymom}
\langle E^p \rangle_c= {d N \over 2 \ell_m} \int_0^{\ell_m} T^p (\ell)
\textrm{d} \ell \,\,\,.
\end{equation}
In figure \ref{figenpdf} the Inverse Fourier Transform of
(\ref{charfunen}) is compared with the function $\Pi (E)$ previously
defined. The similarity of the two functions is remarkable.
Another important feature that can be checked with this results is the
dependence of the above defined two macroscopic quantities ($N_f$ and
$T_E$) with system size. It is straightforward to see, from (\ref{tgran})
and (\ref{energymom}), that the granular temperature and the total
kinetic energy are respectively an intensive and an extensive variable
if $\ell_m$ is independent from the system size. This is effectively
the case if both the density $\rho=N/V$ and the total height $H$ are
kept constant.   Moreover, for large enough $\ell_m$, 
the integral in equation~\eqref{energymom} becomes size independent:
\begin{equation}
\int_0^{\ell_m} T^p(\ell) \textrm{d}\ell  \sim {T_0^p \over 2 p
  \sqrt{a(\alpha)}} \,\,.
\end{equation}
Thus, the effective temperature $T_E$ defined above becomes a constant
proportional to the temperature of the wall, while the parameter $N_f$ still
depends on the system size:
\begin{equation}
N_f \sim  {1 \over \sqrt{a(\alpha)}} \, {d N \over  \ell_m}\,,
\;\;\;\;\;\;\;\;\;\;\;\;\;\;\;\;\;\;\;\;
T_E\sim {T_0 \over 2}.
\end{equation}
Numerical simulations show that $T_E$ effectively remains a constant for
large systems, and under several procedures of box size increase. The behavior
of $N_f$ is determined by the maximum of the integrated density
$\ell_m$. For a square cell at constant density one finds $\ell_m
\propto \sqrt{N}$ , so that $N_f \propto \sqrt{N}$, which is not far
from $N^{0.4}$ observed in \cite{aumaitre04}. Moreover, if only the
height $H$ of the cell is increased, $\ell_m$ is proportional to $N$,
and $N_f$ becomes constant. All those features are in agreement with
the numerical observations in \cite{aumaitre04}.  The above results
clearly show that a rough calculation, which takes into account only
the inhomogeneities of the system, is able to quantitatively describe
the behavior of the fluctuations of the total kinetic energy of a
vibrated granular gas.  In some cases the energy pdf can be
approximated with a gamma distribution, which is the standard
distribution for the energy pdf in the canonical equilibrium.
Nevertheless there are strong deviations from the equilibrium theory
of fluctuations, since the two parameters of the gamma distribution
(i.e. the temperature and the number of degrees of freedom) are not
the granular temperature neither the number of degrees of freedom.
Another important remark is that correlations, and in particular
contributions from the two points distribution function, do not play a
primary role to explain those deviations from the equilibrium theory
of fluctuations. In order to characterize corrections arising from the
two particles velocity pdf, one should measure energy fluctuations at
a given height $x$ from the vibrating wall. As already noted in
\cite{aumaitre04} this task is very hard, since the available
statistic become very poor.  Nevertheless an effective way to quantify
those fluctuations is to look at homogeneous systems, where
contributions coming from the inhomogeneities vanish. 
With this objective in mind,
we will be interested in the following in granular gases heated by an
homogeneous and isotropic driving.

\subsection{The homogeneously driven case}

In this section we will present some numerical results concerning the energy
fluctuations in a dilute gas driven by the stochastic thermostat presented in
section~\ref{randomlydriven}~\cite{viscoepj}. When the system reaches a stationary state, the
dissipated energy is compensated by the energy injected by the thermostat, and
the temperature fluctuates around its mean value.
Here we are interested in the fluctuations of the total energy measured by the
quantity
\begin{equation}
\sigma_E^2=N {\langle E^2(t) \rangle - \langle E(t) \rangle^2 \over \langle
  E(t) \rangle^2}\,\,.  
\end{equation}
Note that $\sigma_E^2\equiv 2N/N_f$. Brey {\it et al.} have computed, by means
of kinetic equations, an analytical expression for $\sigma_E^2$ in the
homogeneous cooling state, which is equivalent to the so-called Gaussian
deterministic thermostat.  One of the main differences of
this stochastic thermostat with a deterministic one, is found in the elastic
limit. On the one hand, for the cooling state, when the restitution
coefficient tends to 1, the conservation of energy imposes that the energy pdf
is a Delta function, and the quantity $\sigma_E$ goes to $0$. On the other
hand, with the stochastic thermostat, if the elastic limit is taken keeping
the temperature constant, the strength of the white noise will tend to zero,
but it will still play a role in the velocity correlation function.

\begin{figure}
  \centering \includegraphics[clip=true,width=0.55
  \textwidth]{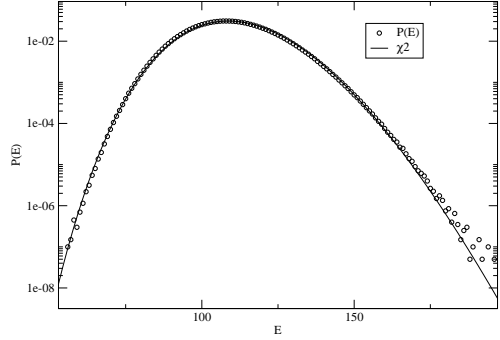}
  \caption{\label{figenpdfdriv} Energy pdf (dots) from DSMC simulations with a 
    restitution coefficient $\alpha = 0.5$ and $N=100$ particles for a system
    driven with the stochastic thermostat. The solid line shows a gamma
    distribution with same mean and same variance.}
\end{figure}

We performed DSMC simulations to measure the energy pdf of such a
system. A plot of this pdf is shown in figure \ref{figenpdfdriv}, and
it is close to a $\chi^2$-distribution with
same mean and same variance. Nevertheless the number of degrees of
freedom of this $\chi^2$-distribution is lower than the true number of
degrees of freedom (i.e. $(N-1) \times d $).  This effect may arise
from two separated causes: the non-gaussianity of the velocity pdf,
and the presence of correlations between the velocities. This feature
also suggests that a calculation of the energy pdf with the hypothesis
of uncorrelated velocities (but non-Gaussian) could explain at least a
part of this non-trivial effect. In order to quantify these
contributions we will consider that the velocity pdf is well described
by a Gaussian multiplied by the second Sonine polynomial:
\begin{equation}
\label{pdfsonine}
f(v)=\frac{e^{-\frac{v^2}{2 T}}}{(2 \pi T)^{d/2}} \left(1+a_2 S_2 \left(\frac{v^2}{2 T} \right) \right)\,\,,
\end{equation}
where $a_2$ is given by expression (\ref{a2}).

\begin{figure}
  \centering 
  \includegraphics[clip=true,width=0.55
  \textwidth]{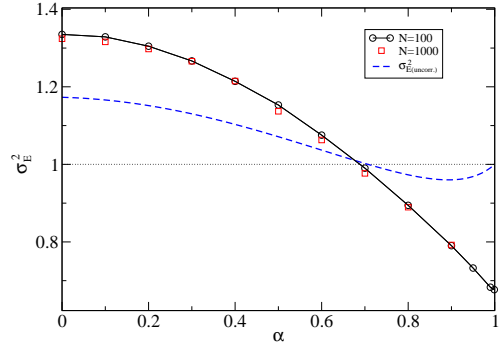}
\caption{\label{ealphadriven} Plot of $\sigma^2_E$ versus the restitution
  coefficient $\alpha$ for $N=100$ ($\bigcirc$) and $N=1000$
  (\textcolor{red}{$\square$}) particles. The result of the calculation
  assuming uncorrelated velocities (\ref{sigmauncorr}) is shown by the dashed
  line.}
\end{figure}

The calculation of the pdf of the sum
of the square of $N$ variables distributed following (\ref{pdfsonine})
is straightforward. The characteristic function of the energy pdf is:
\begin{equation}
{\widetilde P}_N(k)= {1 \over (1-\im k T)^{Nd \over 2}} \
\left(1+{d (d+2) \over 8} a_2
\left({1 \over (1- \im k T)^2}- {2 \over (1-\im k T)} +1 \right) \right)^N
\end{equation}
where $N$ is the number of particles of the system.  This yields:
\begin{equation}
\langle E \rangle = {d \over 2} N T\,\,,
\;\;\;\;\;\;\;\;\;\;\;\;\;\;\;\;\;\;\;\;
\langle E^2 \rangle - \langle E \rangle^2={d \over 2} N T^2 \left(1+{d+2 \over
    2} a_2 \right) \,\,.
\end{equation}
It is now possible to have an explicit expression for the energy fluctuations:
\begin{equation}
\label{sigmauncorr}
\sigma^2_{E_{(uncorr.)}}={2 \over d} \left(1+{d+2 \over 2} a_2 \right) \,\,.
\end{equation} 
In figure \ref{ealphadriven} this result is compared with the result of DSMC
simulations, performed for several values of the restitution coefficient
$\alpha$ and for two different values for the number of particles $N$. The
disagreement between the uncorrelated calculation and the simulations is a
clear sign of the correlations induced by the inelasticity of the system. One
can note that the fluctuations increase when the restitution coefficient
decreases. One can also see that there is a value of the restitution
coefficient $\alpha$ around $1/\sqrt{2}$, that is when the approximate
expression of $a_2$ vanishes, for which $\sigma_E^2$ is exactly $1 \equiv
2/d$, as for a gas in the canonical equilibrium (velocities are then
uncorrelated).

\begin{figure}
  \centering
  \includegraphics[clip=true,width=0.55 \textwidth]{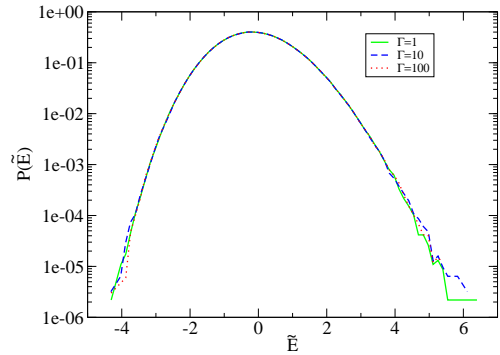}
  \caption{\label{allrescaled} Plot of the pdf of the rescaled energy
   ${\widetilde E}$ for a 
    restitution coefficient $\alpha=0.5$ and for $N=100$ for several values of
    the strength of the noise $\Gamma$.}
\end{figure}

We now turn to the dependence of $\sigma^2_E$ on 
the strength of the white noise $\Gamma$. It is useful, for
this purpose, to introduce a rescaled, dimensionless energy
\begin{equation}
{\widetilde E}={E - \langle E \rangle \over \sqrt{\langle E^2 \rangle -
    \langle E \rangle^2}} \,\,.
\end{equation} 
We have plotted in figure \ref{allrescaled} this rescaled energy pdf for a
system of $N=100$ particles with a restitution coefficient $\alpha=0.5$ and
for several values of the strength of the white noise $\Gamma$. One can see
how all the pdfs collapse into a unique distribution. The role of the noise's
strength is thus only to set the temperature (or mean kinetic energy) scale.
Besides, relative energy fluctuations depend only on $\alpha$ and $N$.
Moreover, since $\sigma_E^2$ does not depend on the number of particles $N$
(for $N$ large enough), the central limit theroem applies, and hence
$P({\tilde E})$ is a Gaussian in the thermodynamic ($N \to \infty$) limit.
In conclusion we have shown that randomly driven granular gases display non
trivial fluctuations, because of the correlations induced by the inelasticity.
Two different kinds of correlations contribute to this behavior of the
fluctuations. First, the non-Gaussianity of the velocity pdf, which simply
tells that the Euclidean components of the velocity of each particle are
correlated one to each other. Second, a contribution from the two particles
velocity pdf, which does not factorize exactly as a product of two
one-particle distributions.  It must be pointed out, however, that these
correlations do not invalidate the Boltzmann equation. As already noted in
\cite{ernst81,brey04a,maxfluct}, the two points correlation function $g_2(\bv_1,
\bv_2)$, which is defined by:
\begin{equation}
 g_2(\bv_1, \bv_2)=f^{(2)}(\bv_1, \bv_2)-f(\bv_1)f(\bv_2)\,\,,
\end{equation}
where $f^{(2)} $ is the two points distribution, is of higher order in
the density expansion (roughly speaking ${\cal O} \left(g(\bv_1,
\bv_2) \right) \sim {\cal O} \left(f(\bv_1)f(\bv_2) \right)/N$).  This
is confirmed by the numerical observations, since when the number of
particles increases, the energy pdf tends be closer and closer to a
Gaussian.  Spatial correlations, which can be at work in homogeneously
driven granular gases, at higher densities, and which have been
negleted here (assuming spatial homogeneity) can also play a relevant
role in fluctuations~\cite{fdtvulpio}.


\section{A large deviation theory for the injected power fluctuations in the homogeneous driven granular gas}
\label{ipfluct}

From now on we turn our attention to the fluctuations of another global
quantity, i.e. the power injected into the system by the external source of
energy.  In particular in this section our main goal is to obtain a kinetic
equation able to describe the behavior of the large deviations function of
the time integrated injected power~\cite{visco05,visco05b} in a randomly driven
gas (cf section~\ref{randomlydriven}). The latter quantity is the total work
${\cal W}$ provided by the thermostat over a time interval $[0,t]$:
\begin{equation}
{\cal W}(t) = \int_0^t \dd t \sum_i \bF^{th}_i \cdot \bv_i\,.
\end{equation} 
Our interest goes to the distribution of ${\cal W}(t)$, denoted by $P({\cal
  W},t)$, and to its associated large deviation function $\pi_{\infty}(w)$
defined for the reduced variable $w={\cal W}/t$ (${\cal W}(t)$ being extensive
in time):
\begin{equation}
\pi_{\infty}(w)=\lim_{t \to \infty} \pi_t(w) \,\,\,,
\pi_t(w) =\frac{1}{t}\log P({\cal W}=wt,t)\,.
\end{equation}
We introduce $\rho(\Gamma_N,{\cal W},t)$ the probability that the system is in
state $\Gamma_N$ at time $t$ with ${\cal W}(t)={\cal W}$. The function we want
to calculate is
\begin{equation}
P({\cal W},t)=\int\dd\Gamma_N \rho(\Gamma_N,{\cal W},t).
\end{equation}
We shall focus on the generating function of the phase space density
\begin{equation}
\hat{\rho}(\Gamma_N,\lambda,t)=\int\dd {\cal W}\ee^{-\lambda 
{\cal W}}\rho(\Gamma_N,{\cal W},t)
\end{equation}
and on the large deviation function of
\begin{equation}
\hat{P}(\lambda,t)=\int\dd
{\cal W}\ee^{-\lambda {\cal W}}P({\cal W},t)=
\int\dd\Gamma_N\hat{\rho}(\Gamma_N,\lambda,t)
\end{equation}
which we define as
\begin{equation}
\mu(\lambda)=\lim_{t \to \infty} \frac{1}{t}\log \hat{P}(\lambda,t)\,\,.
\end{equation}
Note that $\mu(\lambda)$ is the generating function of the cumulants of ${\cal
  W}$, namely
\begin{equation}
\label{cumulants}
\lim_{t \to \infty} \frac{\langle {\cal W}^{n}\rangle_c}{t}=
(-1)^n \left.\frac {\dd^n\mu(\lambda)}{\dd\lambda^n}\right|_{\lambda=0} 
\end{equation}
Moreover $\pi_{\infty}(w)$ can be obtained from $\mu(\lambda)$ by means of a
Legendre transform, i.e. $\pi_{\infty}(w)=\mu(\lambda_*)+\lambda_*w$ with
$\lambda_*$ such that $\mu'(\lambda_*)=-w$.

The observable ${\cal W}$ is non-stationary but it is Markovian, hence a
generalized Liouville equation for the extended phase-space density
$\rho(\Gamma_N,\cW,t)$ can be written. It varies in time under the combined
effect of the inelastic collisions (which do not alter $\cW$)
and of the random kicks:
\begin{equation}
\p_t\rho=\p_t\rho\Big|_{\text{collisions}}+\p_t\rho\Big|_{\text{kicks}}
\end{equation}
Considering that the thermostat acts independently on each
particle, it can be shown that
\begin{equation}
\p_t\hat{\rho}\Big|_{\text{kicks}}=
\sum_i\Big[ \Gamma(\Delta_{\bv_i}+2\lambda \Gamma
\bv_i \cdot \p_{\bv_i}+\Gamma(d\lambda+\lambda^2 v_i^2)\Big]\hat{\rho}
\end{equation}
This additional piece is linear in $\hat{\rho}$ just as the collision part is.
The large time behavior of $\hat{\rho}$ is governed by the largest eigenvalue
$\mu(\lambda)$ of the evolution operator of $\hat{\rho}$. In the large time
limit, we thus expect that
\begin{equation}
\hat{\rho}(\Gamma_{N},\lambda,t)\simeq C(\lambda) \ee^{\mu(\lambda) t}\tilde
{\rho}(\Gamma_N,\lambda),
\end{equation}
where $\tilde{\rho}(\Gamma_N,\lambda)$ is the eigenfunction associated to
$\mu$, and $C(\lambda)$ is such that $\tilde{\rho}(\Gamma_N,\lambda)$ is
normalized to unity.  We then introduce
\begin{equation}\hat{f}^{(k)}(v_1,\dots,v_k,\lambda,t)=\int\dd\Gamma_{N-k}\hat{\rho},
\end{equation}
where $\int\dd\Gamma_{N-k}$ means an integration over $N-k$ particles, we have that
\begin{equation}\label{base}
\p_t\hat{f}^{(1)}(v,\lambda,t)=\Gamma\Delta_{\bv}\hat{f}+2\lambda \Gamma
\p_{\bv} \cdot \bv \hat{f}+\Gamma(\lambda^2 v^2-d \lambda)\hat{f}+\hat{J}
\end{equation}
with $\hat{J}=\int\dd {\cal W}\ee^{-\lambda {\cal W}}J$ the Laplace transform
of the collision integral in which $f(v,{\cal W},t)$ now plays the role of the
velocity distribution. Quite unexpectedly the above equation has a straight
physical interpretation: consider a many particle system where a noise of
strength $\Gamma$ and a viscous friction-like force $\bF=-2 \lambda \Gamma
\bv$ act independently on each particle, and where the particles interact by
inelastic collisions. Consider then that the particles annihilate/branch
(depending on the sign of $\lambda$) at constant rate $d \lambda \Gamma$, and
branch with a rate proportional to $\lambda^2 v^2 \Gamma$. Then, the equation
governing the evolution of the one particle velocity distribution of such a
system is exactly the equation (\ref{base}), where $\lambda$ is a parameter
tuning the strength of the external fields. Moreover, in spite of there being
no {\it a priori} reason for that, $\tilde \rho$, as well as $\tilde f=\int
\dd \Gamma_{N-1} {\tilde \rho}$, can be interpreted as probability density
functions.

The one and two-point functions ${f}^{(1)}(v,{\cal W},t)$ and
$f^{(2)}(v_1,v_2,{\cal W},t)$ that enter the expression of $J$ are expected to
verify, at large times,
\begin{equation}
\hat{f}^{(1)}(v_1,\lambda,t)= C(\lambda) \ee^{\mu t}
\tilde{f}^{(1)}(v_1,\lambda),
\end{equation}
and
\begin{equation}
\hat{f}^{(2)}(v_1,v_2,\lambda,t)=C(\lambda) \ee^{\mu t}
\tilde{f}^{(2)}(v_1,v_2,\lambda),
\end{equation}
where both $\ft^{(1)}$ and $\ft^{(2)}$ are normalized to unity.  We perform
the following molecular-chaos-like assumption:
\begin{equation}\label{truncation}
\tilde{f}^{(2)}(v_1,v_2,\lambda)\simeq\tilde{f}^{(1)}(v_1,\lambda) 
\tilde{f}^{(1)}(v_2,\lambda)
\end{equation}
which does have a definite physical interpretation in the language of the
inelastic hard-spheres with fictitious dynamics (viscous friction, velocity
dependent branching/annihilation) described in the above paragraph.  Then we
get that
\begin{equation}\label{lambdaBoltzmann}\begin{split}
    \mu\tilde{f}(v,\lambda)=&\Gamma\Delta_{\bv}\tilde{f}+2\lambda \Gamma
    \bv\cdot\p_{\bv}\tilde{f}+\Gamma(d\lambda+\lambda^2 v^2)\tilde{f}\\&+
    \frac{1}{\ell}\int_{\bv_{12}\cdot\bsig>0}\dd v_2 \dd\bsig
    \bv_{12}\cdot\bsig
    \left[\alpha^{-2}\tilde{f}(v_1^{**},\lambda)\tilde{f}(v_2^{**},\lambda)
      -\tilde{f}(v_1,\lambda)\tilde{f}(v_2,\lambda)\right]
\end{split}
\end{equation}
where we have now omitted the superscript $(1)$ denoting the one-point
function. The $\lambda=0$ limiting case yields the usual Boltzmann equation,
since in this case a stationary solution exists, and hence $\mu (\lambda=0)
=0$. The boundary condition to the evolution equation above is thus:
\begin{equation}
\tilde{f}(\bv,\lambda=0)=f_{st}(\bv)
\end{equation}
with $f_{st}(\bv)$ the stationary velocity pdf (cf. Eq.(\ref{stationarypdf})).

\subsection{The cumulants}

\begin{figure}
\begin{minipage}[t]{.46\linewidth}
  \includegraphics[width=1 \textwidth,clip=true]{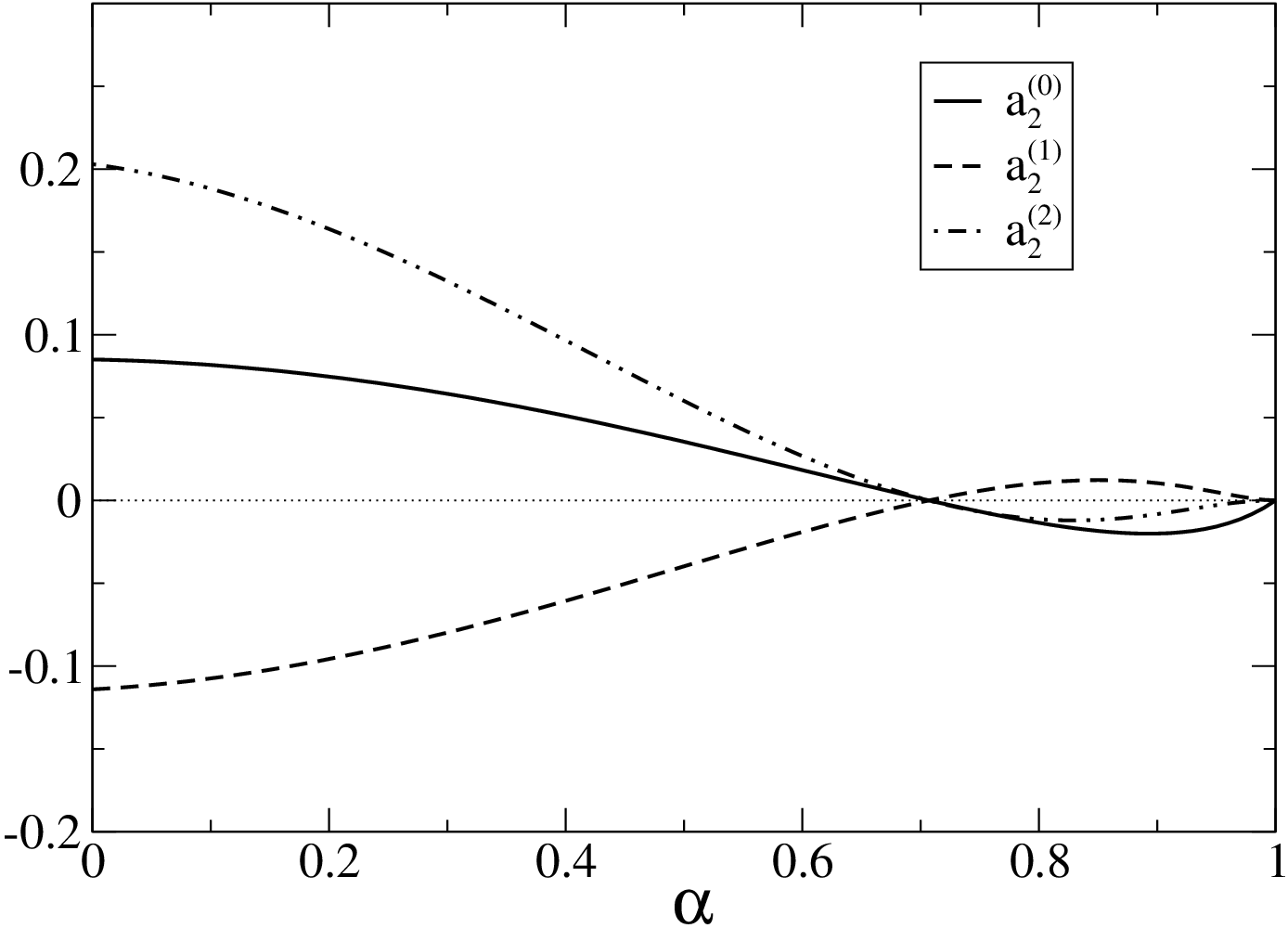}
\caption{\label{a2plot}
$a_2^{(0)}$, $a_2^{(1)}$ and $a_2^{(2)}$ versus $\alpha$ for
$d=2$.}
\end{minipage}
\hfill
\begin{minipage}[t]{.46\linewidth}
  \includegraphics[width=1 \textwidth, clip=true]{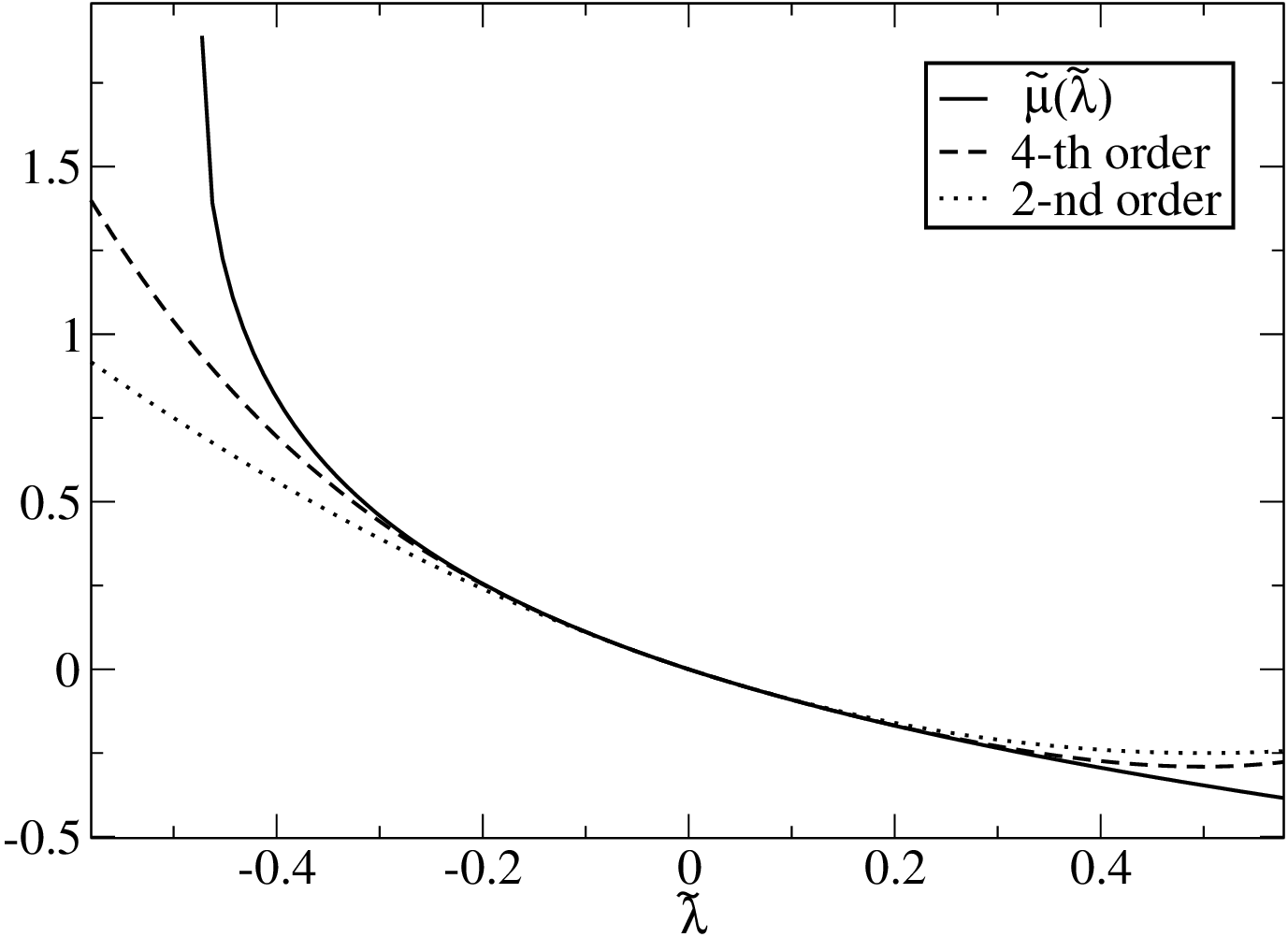}
\caption{\label{plotmunoappr}The solid line shows $\mut$ in the
  limit $d \to \infty$. The dashed line is $\mut$ at fourth order in $\lambt$
  from (\ref{solsys1}) for $d=2$ and $\alpha=0.5$. Finally the dotted line
  shows the same quantity calculated with a truncation at second order in
  $\lambda$, which would satisfy the G-C relation. }
\end{minipage}
\end{figure}

Here we find an approximated expression of $\mu(\lambda)$ solving a
system of equations obtained projecting (\ref{lambdaBoltzmann}) on the first
velocity moments.  First we shall define a dimensionless velocity ${\bf
  c}={\bf v}/v_0(\lambda)$, where $v_0(\lambda)$ plays the role of a thermal
velocity:
\begin{equation}
\label{v02lambda}
v_0^2(\lambda)=2 T (\lambda) = {2 \over d} \int \textrm{d}{\bf v}  \, v^2 \, 
{\tilde f}(v,\lambda)\,.
\end{equation}
Then, defining the function $f(\bc,\lambda)=v_0(\lambda){\tilde f}(\bv,
\lambda)$, and its related moments of order $n$
\begin{equation}
\label{remoments}
m_n (\lambda)= \int \textrm{d}{\bf c} \, c^n f(c,\lambda) \,\,\,,
\end{equation}
one obtains the following recursion relation:
\begin{equation}
\label{moments}
(\mu + \Gamma(2n+d) \lambda)m_n=
{\Gamma \over v_0^2} n(n+d-2)m_{n-2}+\Gamma \lambda^2 v_0^2 
m_{n+2}- v_0   \nu_n,
\end{equation}
where 
\begin{equation}
\nu_n=-\int \textrm{d}{\bf c} \, c^n \, J[f,f]\,\,.
\end{equation}
Recalling the definition of the cumulants (\ref{cumulants}), and the
approximated solution for the stationary velocity pdf, it appears natural to
argue that, for $\lambda \sim 0$, the function $f(\bc, \lambda)$ should be
well approximated by:
\begin{equation}
f(c, \lambda)=\phi(c)\left(1+a_1(\lambda) S_1 \left(c^2 \right)+ a_2(\lambda) 
S_2 \left(c^2 \right) \right) + {\cal O}(a_3)\, \,,
\end{equation}
where $\phi(c)=\pi^{-d/2} \exp (-c^2)$ is the Gaussian distribution.  Even in
this case, from the relation (\ref{coefficients}) and from the definition
(\ref{v02lambda}), the coefficient $a_1$ is found to be 0.  The method
consists in taking the equation (\ref{moments}) for $n=0$, 2 and 4 in order to
find an explicit expression of $\mu$, $v_0$, and $a_2$ in the limit $\lambda
\to 0$.  The quantities $\nu_2$ and $\nu_4$ have been calculated at the first
order in $a_2$ \cite{noije98c}, and their explicit expressions are:
\begin{equation}
\nu_2={(1-\alpha^2) \over 2 \ell} {\Omega_d \over \sqrt{2 \pi}} 
\left\{ 1 + {3 \over 16} a_2 \right\}= {d \Gamma \over \sqrt{2 T_0^3}} 
\left\{ 1 + {3    \over 16} a_2 \right\} \,, 
\end{equation}
and
\begin{equation}
 \nu_4= {d \Gamma \over \sqrt{2 T_0^3}} \left\{ T_1 + a_2 T_2 \right \},
\end{equation}
with
\begin{align}
  T_1 &=d+{3 \over 2} + \alpha^2 \\
  T_2 &={3 \over 32} (10 \, d + 39 + 10 \, \alpha^2) +{ (d-1) \over (1-
    \alpha)}\,,
\end{align}
where $T_0=\bigg({2 d \Gamma \ell \sqrt{\pi} \over (1-\alpha^2)
  \Omega_d}\bigg)^{2/3}$ is the the granular temperature obtained averaging
over Gaussian velocity pdfs (i.e. the zero-th order of Sonine expansion). The
expression of the first moments $m_n$ is:
\begin{subequations}
\begin{gather}
m_0=1 \\ m_2=d/2 \\ m_4=\frac{\left( 1 + a_2 \right) \,d\,
    \left( 2 + d \right) }{4} \\
m_6=\frac{ \left( 1 + 3\, a_2 \right) \,d\,
      \left( 2 + d \right) \,\left( 4 + d \right)  }
    {8}
\end{gather}
\end{subequations}
With the help of the above defined temperature scale $T_0$, we 
introduce some dimensionless variables:
\begin{equation}
\begin{split}
  {\tilde \mu} = \mu {T_0 \over d \Gamma}\,, \quad \quad \quad & \quad \quad
  \quad {\tilde
    \lambda}  = \lambda T_0 \,,\\
  {\tilde v}_0^2 = {v_0^2 \over 2 T_0}\,,\quad \quad \quad & \quad \quad \quad
  {\tilde \nu}_p   = {\sqrt{2 \, T_0^3} \over \Gamma} \nu_p\,.\\
\end{split}
\end{equation}
Note that this scaling naturally defines the scales for the other quantities
of interest, namely: 
\begin{equation}
\begin{split}
  \pit_t=\pi_t {T_0 \over d \Gamma}\,,\quad\quad\,\,\wt={w \over
    d\Gamma}\,,\quad\quad\,\,\cWt={\cW \over \langle \cW \rangle}\,\,.
\end{split}
\end{equation}
The expression of the moment equation (\ref{moments}) becomes, for the above
defined dimensionless quantities:
\begin{equation}
\label{recursiverescaled}
\left( \mut d +(2n+d) \lambt \right) m_n = {n(n+d-2) \over 2 \vt^2}m_{n-2}+
2 \vt^2  m_{n+2} - \vt {\tilde \nu}_n \,\,.
\end{equation}
First we solve the above equation for $n=0$, getting the following result:
\begin{equation}
\label{solsys1}
{\tilde \mu ({\tilde \lambda})} =- {\tilde \lambda} + {\tilde \lambda}^2 
{\tilde v}_0^2({\tilde \lambda}).
\end{equation}
Recalling that when $\lambda \to 0$ one has $v_0^2 = 2 T_g + {\cal
  O}(\lambda)$, it is important to note that if we restrict our
analysis to the Gaussian approximation for $P({\cal W},t)$, that is if we
truncate $\mu (\lambda)$ to order $\lambda^2$, eq. (\ref{solsys1}) will
read:
\begin{equation}
\label{secondorder}
{\mu \over d \Gamma} = \lambda (\lambda T_g -1)\,\,.
\end{equation}
Then we see that indeed
\begin{equation}
\label{symGC}
\mu(\lambda)=\mu\left({1 \over T_g}-\lambda\right)\,\,,
\end{equation}
which means that $\pi_{\infty}(w)=\text{max}_{\lambda} \{\mu(\lambda)+\lambda
w\}$ verifies
\begin{equation} \label{pi_secondorder}
\pi_{\infty}(w)-\pi_{\infty}(-w)={w  \over T_g}\,\,.
\end{equation}
However, the nontrivial functions $m_n(\lambda)$ will break the
property (\ref{symGC}), as we shall explicitly show later.  In order
to characterize more precisely the dependence of $\mut$ upon $\lambt$
for small values of $\lambt$ , it is useful to expand ${\tilde v}_0^2$
and $a_2$ in powers of ${\tilde \lambda}$:
\begin{subequations}
\label{lambdaexpansion}
\begin{gather}
  {\tilde v}_0^2({\tilde \lambda}) ={\tilde v}_0^{2^{(0)}}+{\tilde
    \lambda} {\tilde v}_0^{2^{(1)}}+{\tilde \lambda}^2 {\tilde v}_0^{2^{(2)}}
 + \cO (\lambt^3)\\
  a_2(\tilde{\lambda}) =a_2^{(0)}+{\tilde \lambda} a_2^{(1)}+{\tilde
    \lambda}^2 a_2^{(2)}+ \cO (\lambt^3)
\end{gather}
\end{subequations}
In this way we can find ${\tilde v}_0^{2^{(i)}} \left(a_2^{(i)}
\right)$ solving equation (\ref{recursiverescaled}) for $n=2$:
\begin{equation}
{\tilde v}_0^{2^{(0)}}= \left(1- {a_2^{(0)}\over 8} \right)\,\,,
\end{equation}

\begin{equation}
{\tilde v}_0^{2^{(1)}}=-{4 \over 3} + {a_2^{(0)} \over 3} -
{a_2^{(1)} \over 8}\,\,,
\end{equation}

\begin{equation}
{\tilde v}_0^{2^{(2)}}= 2 - a_2^{(0)} \left({1 \over 12} + {d \over 3} \right)
+ {a_2^{(1)} \over 3}- {a_2^{(2)} \over 8} \,\,.
\end{equation}
Then we substitute ${\tilde v}_0^2 ({\tilde \lambda})$ in the third equation
and expand it in powers of ${\tilde \lambda}$ to find the expression of
$a_2^{(i)}(\alpha)$. Note that one has also to expand in powers of $a_2$ and
keep only the linear terms in order to be coherent with the ${\tilde \nu}_p$
calculations.  We find the following expressions, which are plotted in Fig.
\ref{a2plot}:

\begin{equation} \label{a20}
a_2^{(0)}=\frac{4\,\left( 1 - \Mvariable{\alpha} \right) \,
    \left( 1 - 2\,{\Mvariable{\alpha}}^2 \right) }
  {19 + 14\,d - 3\,\Mvariable{\alpha}\,
    ( 9 + 2d ) + 6 ( 1 - \Mvariable{\alpha} ) \,
       \alpha^2  }
\end{equation}

\begin{equation} \label{a21}
a_2^{(1)}= -\frac{4\,{\left( 1 - \Mvariable{\alpha} \right) }^2\,
    \left( -1 + 2\,{\Mvariable{\alpha}}^2 \right) \,
    \left( 31 + 2\,{\Mvariable{\alpha}}^2 + 16\,d \right) }
    {(19 + 14\,d - 3\,\Mvariable{\alpha}\,
    ( 9 + 2d ) + 6 ( 1 - \Mvariable{\alpha} ) \,
       \alpha^2  )^2}
\end{equation}

\begin{equation}
a_2^{(2)}={A(\alpha) \over B(\alpha)}
\end{equation}
with

\begin{equation}
\begin{split}
  A(\alpha)& =16\,{\left( -1 + \Mvariable{\alpha} \right) }^2\,\left( -1 +
    2\,{\Mvariable{\alpha}}^2 \right) \, \{ 906 + \Mvariable{\alpha}\,\left[
    -984 + \Mvariable{\alpha}\,\left( 85 + 3\,\Mvariable{\alpha}\, \left( -19
        + 6\,\left( -1 + \Mvariable{\alpha} \right) \,\Mvariable{\alpha}
      \right) \right) \right] + 985\,d + \\& + \Mvariable{\alpha}\,\left[ -951
    + \Mvariable{\alpha}\, \left( -25 + 3\,\Mvariable{\alpha}\,\left( 7 +
        6\,\left( -1 + \Mvariable{\alpha} \right) \,\Mvariable{\alpha} \right)
    \right) \right] \,d + \left( 269 + 3\,\Mvariable{\alpha}\,\left( -75 +
      2\,\Mvariable{\alpha}\,\left( -7 + 3\,\Mvariable{\alpha} \right) \right)
  \right) \,d^2 \} \, ,
\end{split}
\end{equation}
and
\begin{equation}
B(\alpha)=3\,
    {\left( -19 - 14\,d + 3\,\Mvariable{\alpha}\,\left( 9 + 2\,\left( -1 
+ \Mvariable{\alpha} \right) \,\Mvariable{\alpha} + 2\,d \right)  \right) }^3
\end{equation}
The $v_0^{2^{(0)}}$ expression, as well as the $a_2^{(0)}$ expression, gives
the usual results established for granular gases
\cite{noije98c,montanero00b}.  At this point the computation of the
cumulants becomes straightforward. From relation (\ref{cumulants}) it follows:
\begin{equation}
\lim_{t \to \infty} \frac{\langle \cW^n \rangle_c}{t} = (-1)^n N d \Gamma 
T_0^{n-1} n! \, \, \vt^{2^{(n-2)}}\,\,.
\end{equation}
Moreover, since the $a_2^{(i)}$ corrections are numerically small, the zero-th
order (Gaussian) approximation already gives a good estimate for the
cumulants.  Namely, the first cumulants are , in this approximation:
\begin{equation} \label{4cumulants}
\begin{split}
  \left\langle \cW \right\rangle_c = t N d \Gamma\,\,,\quad \quad \quad & \quad
  \quad \quad
  \left\langle \cW^2 \right\rangle_c= 2 t N d \Gamma T_0\,\,,\\
  \left\langle \cW^3 \right\rangle_c = 8 t N d \Gamma T_0^2\,\,,\quad \quad
  \quad & \quad \quad \quad
  \left\langle \cW^4 \right\rangle_c   =  48  t N d \Gamma T_0^3\,\,.\\
\end{split}
\end{equation}
All the above expansions in powers of $\lambda$, at the second order in Sonine
coefficients (e.g. $a_2$) can be carried out just expanding $v_0$ and $a_2$ in
(\ref{lambdaexpansion}) to higher powers of $\lambda$. Moreover, expanding in
higher order in Sonine coefficient (e.g. $a_3$) remains in principle still
possible, but it will involve a higher number of equations in the hierarchy
(\ref{recursiverescaled}) (e.g. $n=6$), and therefore will need the expression
of higher order collisional moments (e.g. $\nu_6$).

\subsection{The solvable infinite dimension limit}

Strong arguments~\cite{visco05b} can be given showing that in high dimensions
$\ft(v,\lambda)$ is not far from a Gaussian. We are therefore led to consider,
in the limit $d \to \infty$, $\ft(\bv, \lambda)$ to be a Gaussian with a
$\lambda$-dependent second moment. In this situation the dimensionless
function $f$ will read:
\begin{equation}
f(\bc)={e^{-c^2} \over \pi^{d/2}}
\end{equation}
with ${\bf c}={\bf v}/v_0(\lambda)$.  In this context one can solve
equation (\ref{recursiverescaled}) in order to get an explicit expression for
$\mu(\lambda)$.  Solving the system defined by (\ref{recursiverescaled}) for
$n=0$ and $n=2$ gives a unique solution for $\mut(\lambt)$ which verifies the
physical requirement $\mut(0)=0$:

\begin{equation}
\mut (\lambt)=-\lambt + {\lambt^2 \over 2} {\tilde v}_0^2(\lambt)\,,
\end{equation}
with:
\begin{equation}
{\tilde v}_0^2(\lambt)= 
\frac{1 + 4 \,{\lambt}^3}{2\,{\lambt}^4} +
{b_1(\lambt) \over 2}-{1 \over 2}
\left [ -\frac{32}{{\lambt}^2} + \frac{2\,{\left( 1 + 4\,{\lambt}^3 
 \right) }^2}{{\lambt}^8}+b_2(\lambt)-b_3(\lambt)+ 
{b_4(\lambt) \over 4 b_1(\lambt)} \right]^{1 \over 2} \,\,,
\end{equation}
and
\begin{equation}
\begin{split}
b_1(\lambt)= \sqrt{{\lambt}^{-8} 
+ \frac{8}{{\lambt}^5} - 
        b_2(\lambt)+b_3(\lambt)}\,, \quad &
b_2(\lambt)=\frac{16\,{\left( \frac{2}{3} \right) }^{\frac{1}{3}}}
{{\lambt}^3 \,{\left( 9 + {\sqrt{3}}\,{\sqrt{27 + 
256\,{\lambt}^3}} \right) }^{\frac{1}{3}}} \,,\\
b_3(\lambt)=\frac{2\,{\left( \frac{2}{3} \right) }^{\frac{2}{3}}\,
{\left( 9 + {\sqrt{3}}\,{\sqrt{27 + 256\,{\lambt}^3}} 
\right)}^{\frac{1}{3}}}{{\lambt}^4}\,, \quad &
b_4(\lambt)=\frac{256}{{\lambt}^3} - \frac{192\,\left( 1 + 
4\,{\lambt}^3 \right) }{{\lambt}^6} + 
\frac{8 \,{\left( 1 + 4\,{\lambt}^3 \right) }^3}{{\lambt}^{12}}\,.\\
\end{split}
\end{equation}

\begin{figure}
  \centering
\includegraphics[width=0.5 \textwidth, clip=true]{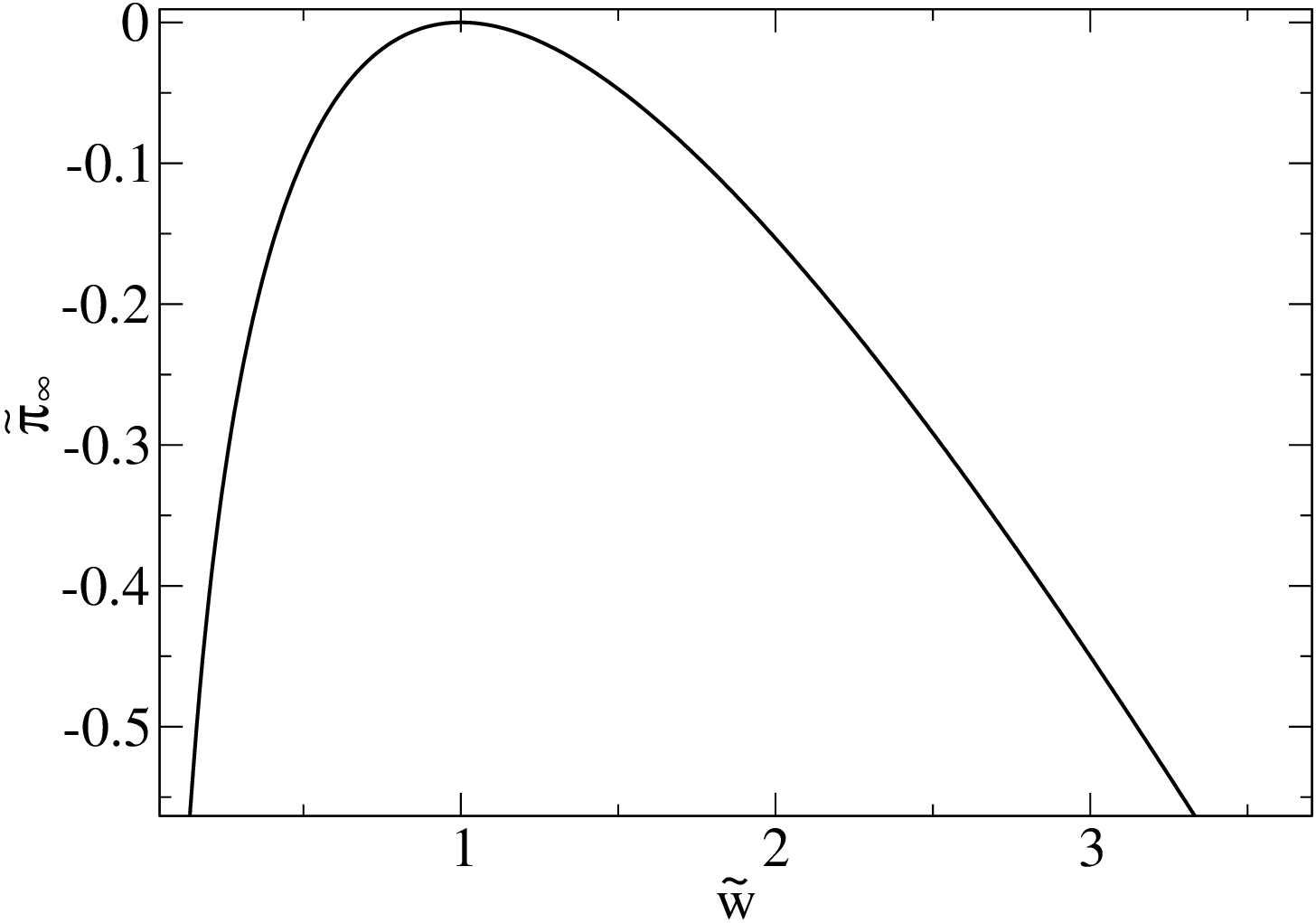} 
\caption{$\pit_{\infty}(\wt)$}
\label{piofw} 
\end{figure}

This expression of the velocity scale reduces to the kinetic temperature for
$\lambda=0$, and decreases monotonically as $\lambda^{-1/2}$ when $\lambda \to
\infty$. This means that in the limit $\lambda \to \infty$ $\ft$ approaches a
Dirac distribution as $\exp(-\lambda v^2/2)$. This feature supports the
intuition that the small ${\cal W}$ events (which are related to the large
values of $\lambda$) are provided by the small velocities. The behavior of
$\mut$ is shown in Fig.  \ref{plotmunoappr}.  The large deviations function
$\mut (\lambdat)$ becomes complex for $\lambdat < -{3 \over 2^{8/3}}$, because
of the terms containing $\sqrt{27 + 256 \lambdat^3}$.  Moreover for large
$\lambdat$ the behavior of this function is $\mut (\lambdat) \sim -
\lambdat^{1 \over 4}$.  In the vicinity of the singularity (i.e.  $\lambdat=
\lambda_0 =-{ 3 \over 2^{8/3}}$) the behavior of the large deviation function
is:
\begin{equation}
\mut (\lambdat) = {3\over 2^{3/2}} - 3^{2/3} 2^{1/6} \sqrt{\lambdat -
    \lambda_0}
+ {\cal O}(\lambdat- \lambda_0)\,\,.
\end{equation}
From the behavior for large $\lambt$ it is possible to recover the left tail
of the large deviation function $\pi_{\infty}$. In general, if $\mu(\lambda)
\sim -\lambda^{\beta}$ for $\lambda \to \infty$, this leads to
$\mu'(\lambda_*)=-\beta \lambda_*^{\beta-1}=-w$. This last relation tells us
that for $\beta<1$ we are recovering the limit $w\to 0^+$, with a behavior of
the large deviation function given by $\pi_{\infty}(w)=\mu(\lambda_*)+
\lambda_* w\sim w^{\beta \over \beta-1}$.  Moreover, from the behavior of
$\mu$ near $\lambda_0$, an analogous calculation provides the right tail of
the large deviation function: $\pi_{\infty}(w) \sim \lambda_0 w$, when $w \to
\infty$.  Finally, in our particular case, the tails are given by
\begin{equation}
\pit_{\infty}(\wt \to 0^+) \sim -\wt^{-1/3}\,\,,
\,\,\pit_{\infty}(\wt\to \infty) \sim -\wt\,\,,
\end{equation}
Note that there is no $w<0$ tail to $\pit_{\infty}$. The graph of the whole function
$\pit_{\infty}(\wt)$ is depicted in Fig. \ref{piofw}.

\section{Fluctuations of injected power at finite times: two examples}
\label{finitetimes}

\subsection{The homogeneous driven gas of inelastic hard disks}

In this section the results of numerical simulations of two models
(inelastic hard spheres and inelastic Maxwell model) are presented with
particular attention to the verification of the Fluctuation Relation for the
injected power. The main requirement to pose the question about the validity
of the Fluctuation Relation is a clean observation of a negative tail in the
pdf of the injected power. This dramatically limits the time $t$ of
integration of $\cW(t)$.  In numerical simulations, as well as in real
experiments, at time larger than a few mean free times the negative tail
disappears. On the other hand, at times of the order of $1$-$3$ mean free
times, the Fluctuation Relation appears to be correctly verified for the
inelastic Hard Spheres model and slightly violated for the inelastic Maxwell
model. The measure of the cumulants, anyway, gives a neat indication of the
fact that the time of convergence of the large deviation function is at least
$10$ times as large and that the true asymptotic is well reproduced by the
theory exposed in this chapter. This theory shows strong arguments against the
validity of a symmetry relation of the Gallavotti-Cohen type for the large
deviations of injected power.

The stationary state of a driven granular gas, modeled by
equation~\eqref{boltz}, under the assumption of Molecular Chaos may be 
studied with a Direct Simulation Monte Carlo technique~\cite{bird94,montanero00b}.  As
a first check of reliability of the algorithm, we have measured the granular
temperature $T_g$ and the first non-zero Sonine coefficient $a_2\equiv
(\langle v^4 \rangle/\langle v^2 \rangle^2-3)/3$.  The measured granular
temperature is always in perfect agreement with the estimate. The measured
$a_2$ coefficient is a highly fluctuating quantity and its average is in very
good agreement with the theoretical estimate.

\begin{figure}
\begin{minipage}[t]{.46\linewidth}
  \includegraphics[width=1 \textwidth,clip=true]{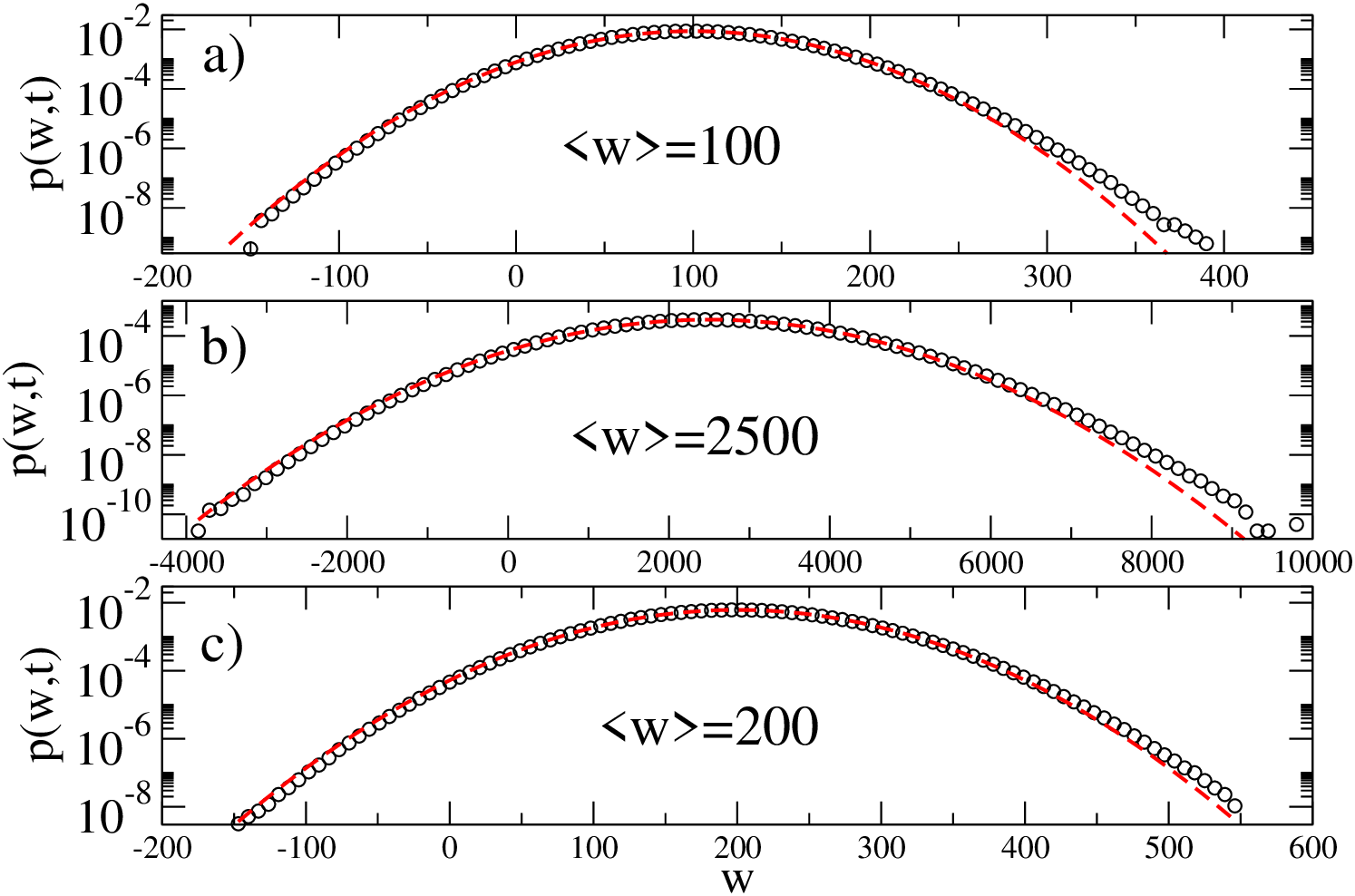}
\caption{Probability density function of the injected power, $p(w,t)\equiv
  tP({\cal W}(t)=wt,t)$ with $t$ equal to $1$ mean free time. In all three
  cases the value of the restitution coefficient is $\alpha=0.9$. Other
  parameters are a) $N=100$, $\Gamma=0.5$; b) N=100, $\Gamma=12.5$; c)
  $N=200$, $\Gamma=0.5$. The dashed line represents a Gaussian with same first
  two cumulants. These distributions have been obtained with $\sim 1.5 \times
  10^9$ independent values of ${\cal W}(t)$.}\label{fig:pdf}
\end{minipage}
\hfill
\begin{minipage}[t]{.46\linewidth}
  \includegraphics[width=1 \textwidth,clip=true]{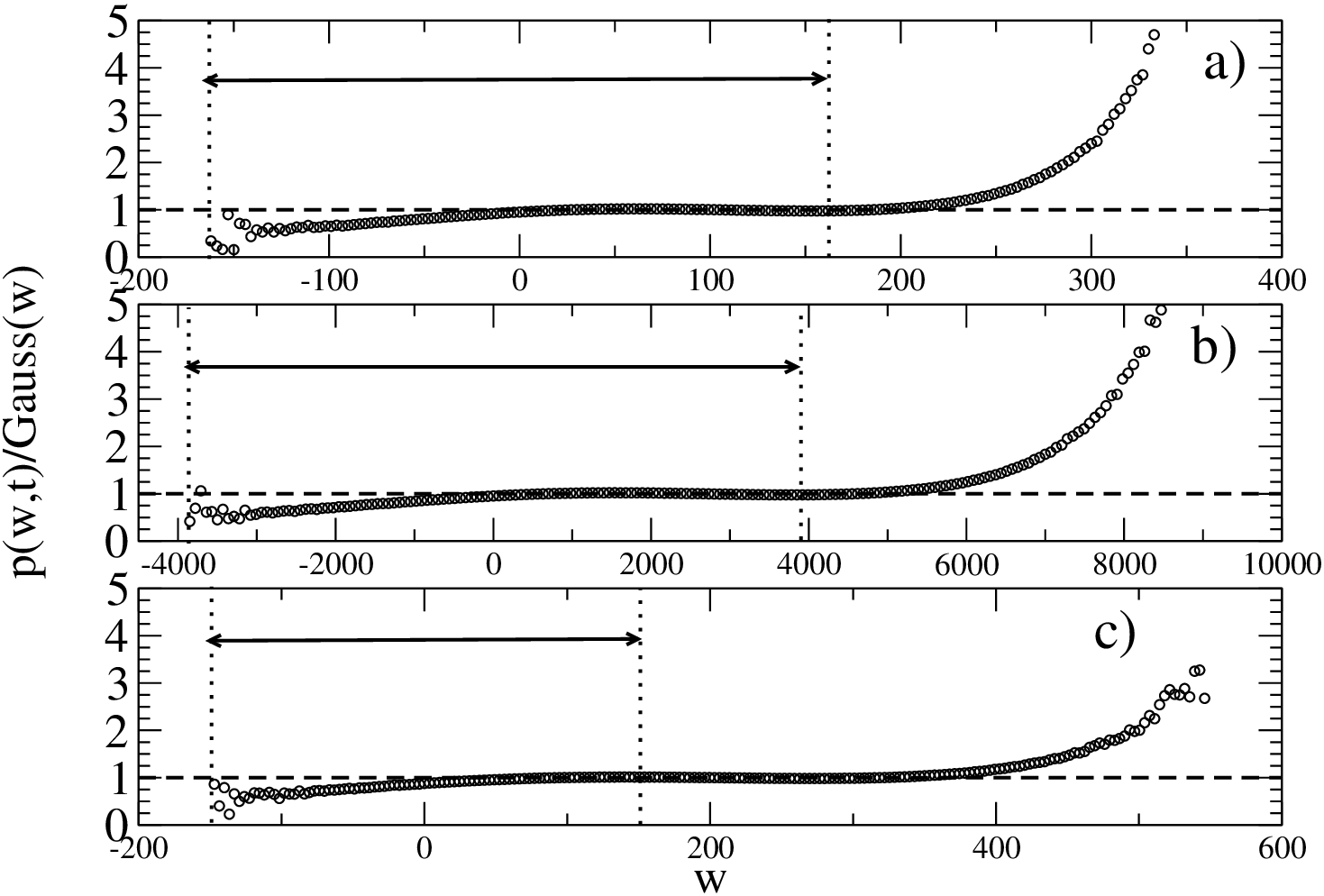}
\caption{Ratio of $P({\cal W},t)$ and a Gaussian with the same first two 
  moments, for the same parameters as in figure~\ref{fig:pdf}: a) corresponds to
  $N=100$, $\Gamma=0.5$, b) to $N=100$, $\Gamma=12.5$ and c) to $N=200$,
  $\Gamma=0.5$.  The range between the vertical dotted lines is the useful one
  for the check of the Gallavotti-Cohen relation. It can be noted that the
  strongest deviations from the Gaussian behavior appear outside of this
  range.}\label{fig:pdfratio}
\end{minipage}
\end{figure}

In figure~\ref{fig:pdf} the probability density functions $p(w,t)\equiv
tP(wt,t)$ (for $t$ equal to $1$ mean free time) for three different
choices of parameters $N,\Gamma$ (at fixed restitution coefficient $\alpha$)
is shown. The values of the first two cumulants of the distribution and their
theoretical values are compared in table~\ref{tab:pdf}, with very good
agreement. In the same table we present also the measure of the third and
fourth cumulants. 

\begin{table}[ht]
\begin{tabular}{|l|l|l|l|l|l|l|l|}
  N   &$\Gamma$  &$\langle {\cal W}(t) \rangle/t$ &$\langle {\cal W}(t)^2 \rangle_c/t$ &$N \Gamma d$
  &$2 N \Gamma d T_g$ &$\langle {\cal W}(t)^3 \rangle_c/t$ &$\langle {\cal W}(t)^4 \rangle_c/t$\\ \hline
  100  &0.5    &100  &20835 &100 &21052 &$6.02779\times10^5$
  &$1.54181\times10^8$\\ \hline
  100  &12.5   &2500 &13019125 &2500 &13157900
  &$9.47684\times10^9$ &$6.12963\times10^{13}$\\ \hline
  200  &0.5    &199.9  &42009  &200  &42120
  &$1.21911\times10^6$ &$3.09634\times10^8$
\end{tabular}
\caption{Rescaled cumulants of the distribution of injected work $P({\cal
  W},t)$, measured with
  $t$ equal to $1$ mean free time for different choices of the parameters.}\label{tab:pdf}
\end{table}

\begin{figure}
\begin{minipage}[t]{.46\linewidth}
  \includegraphics[width=1 \textwidth,clip=true]{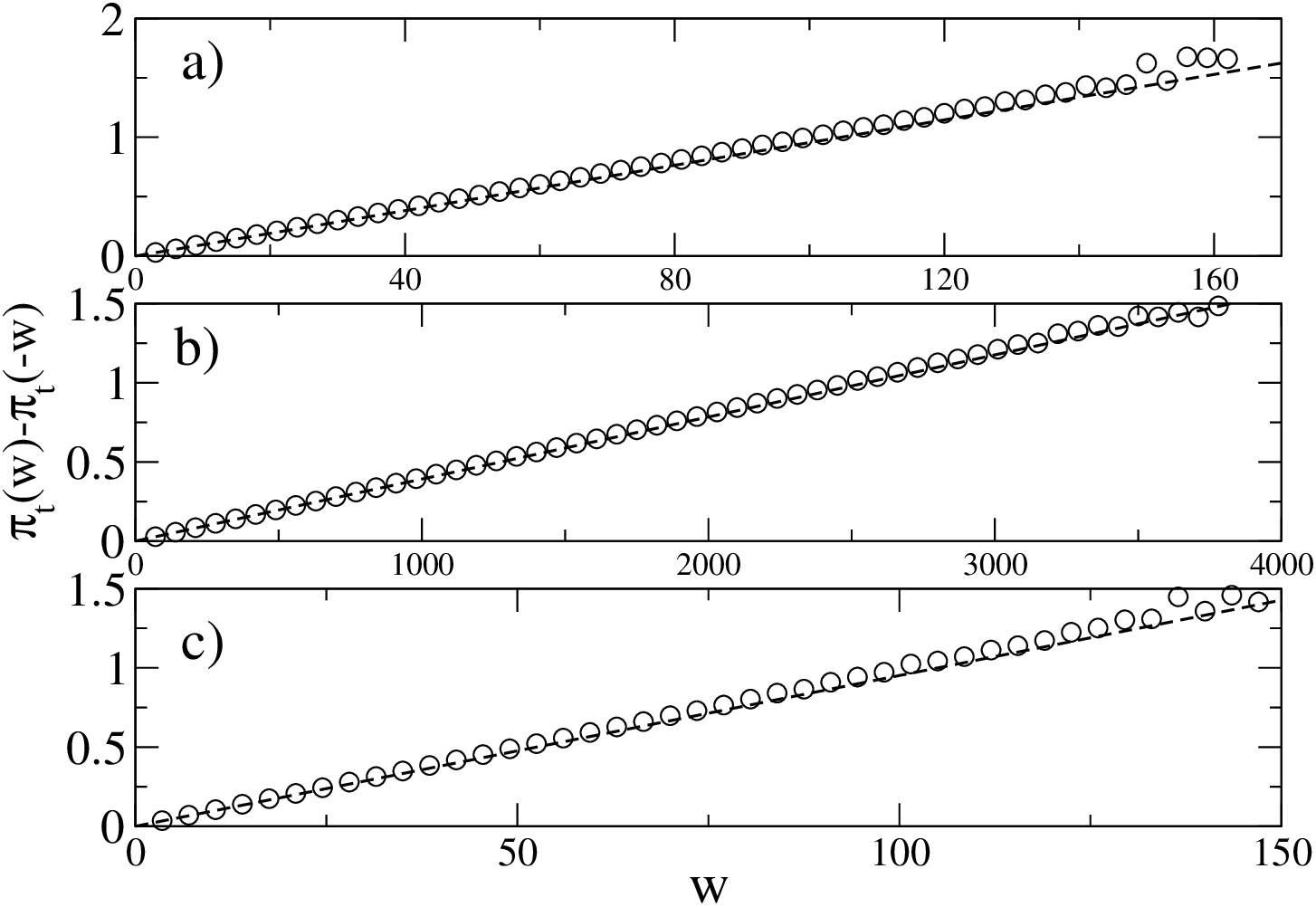}
\caption{Plot of $\pi_t(w)-\pi_t(-w)$ vs $w$, 
for the same choice of the parameters as in
  figure~\ref{fig:pdf}. The values of the slope $\beta_{eff}$ of the fitting
  dashed lines are in table~\ref{tab:gc}}\label{fig:gc}
\end{minipage}
\hfill
\begin{minipage}[t]{.46\linewidth}
  \includegraphics[width=1 \textwidth,clip=true]{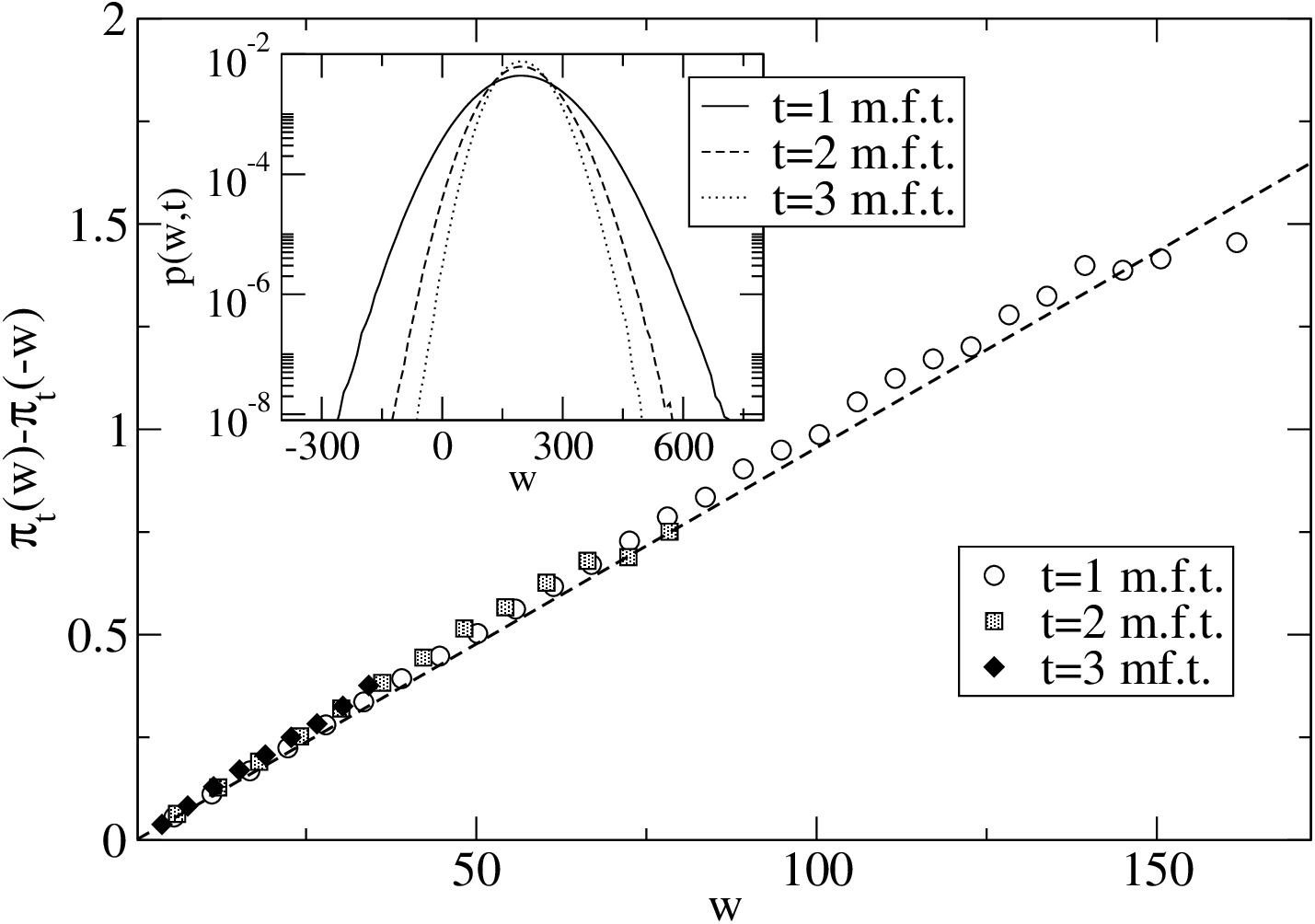}
\caption{Plot of 
  $\pi_t(w)-\pi_t(-w)$ versus $ w$, for the system with $N=100$ and
  $\Gamma=0.5$ for different values of $t$. We recall that in this case
  $\langle w \rangle = 100$. The dashed line has slope $\beta=1/T_g$. In the
  inset the corresponding $p(w,t)$ are shown.}\label{fig:gctau}
\end{minipage}
\end{figure}

The comparison with a Gaussian with same mean value and same variance shows
that the pdf $P({\cal W},t)$ is not exactly a Gaussian. In particular there
are deviations from the Gaussian form in the right (positive) tail. This is
well seen in figure~\ref{fig:pdfratio}. It must be noted that the important
deviations in the right tail arise at values of ${\cal W}(t)$ larger than the
minimum ${\cal W}(t)$ available in the left tail, i.e. they have no influence
in the following plot of figure~\ref{fig:gc} 
regarding the Gallavotti-Cohen symmetry.

In figure~\ref{fig:gc} the Gallavotti-Cohen relation
$\pi_t(w)-\pi_t(-w)=\beta_{eff} w$ is questioned for the same choice of the
parameters. The relation, at this level of resolution and for this value of
the time $t$ ($1$ mean free time), is well satisfied.  Moreover
table~\ref{tab:gc} shows that the value of $\beta_{eff}$ is well approximated
by $\beta=1/T_g$, as expected if the truncation of $\mu(\lambda)$ at the
second order were valid, see eq.~\eqref{pi_secondorder}. In
figure~\ref{fig:gctau} the same relation is checked for different values of
$t$, slightly larger (i.e. up to $t$ equal to $3$ mean free times). No
relevant deviations are observed as $t$ is increased.  Moreover this figure is
important to understand the dramatic consequences that a larger $t$ has on
the ``visibility'' of the Gallavotti-Cohen symmetry: as $t$ is increased,
events with negative integrated power injection become rarer and rarer. This
eventually leads to the vanishing of the left branch of $P({\cal W},t)$.

\begin{table}[ht]
\begin{tabular}{|l|l|l|l|}
  N   &$\Gamma$  &$\beta_{eff}$ &$1/T_g$\\ \hline
  100  &0.5 &0.0100  &0.00955  \\ \hline
  100  &12.5 &0.000402 &0.000382 \\ \hline
  200  &0.5 &0.00995  &0.00952
\end{tabular}
\caption{Factor of proportionality in the ``Gallavotti-Cohen'' relation
  compared with $\beta$.}\label{tab:gc}
\end{table}

\begin{figure}
\begin{minipage}[t]{.46\linewidth}
 \includegraphics[width=1 \textwidth,clip=true]{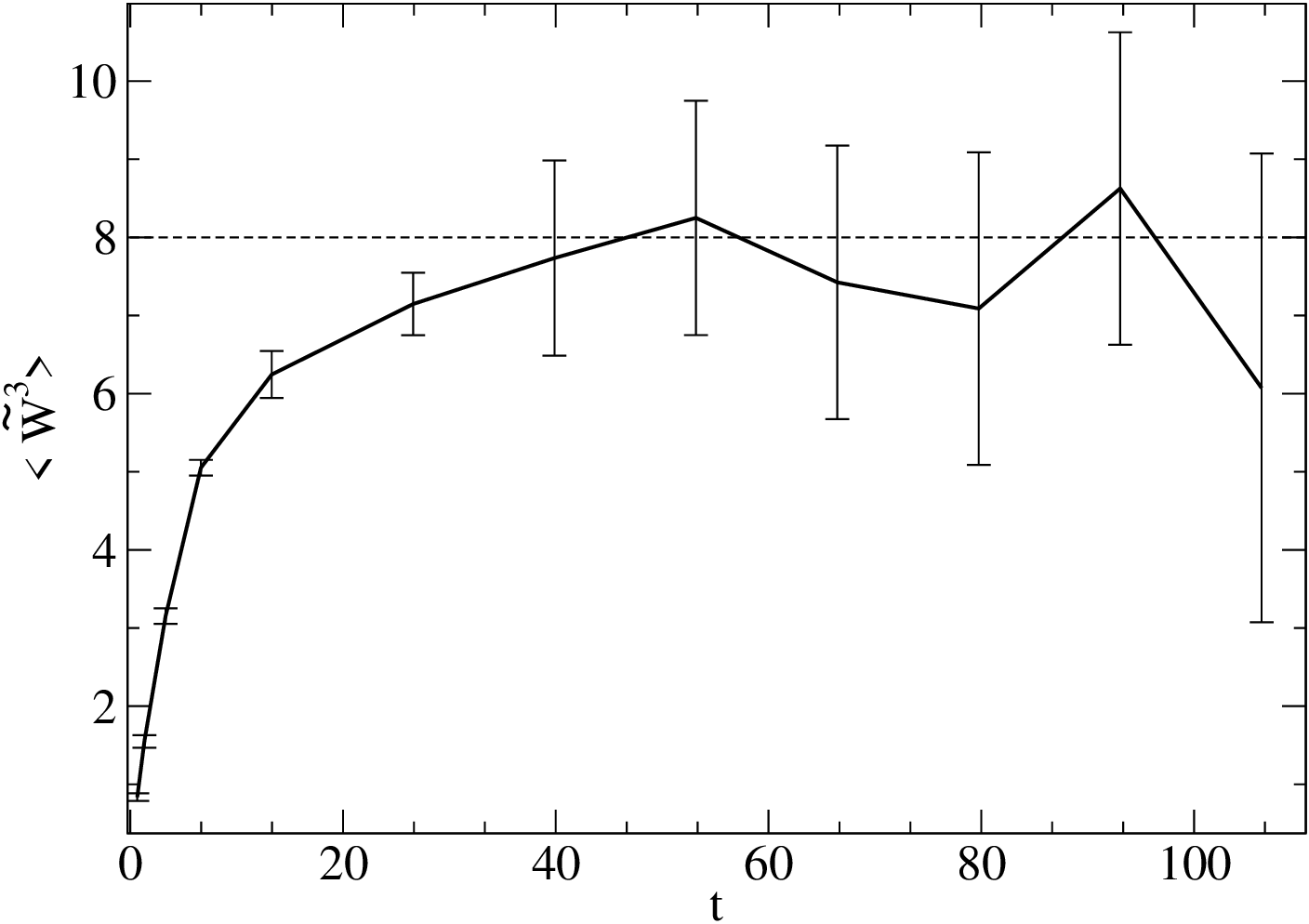}
\caption{\label{thirdcumulant}
  Third cumulant $\cWt=\cW/(\langle \cW \rangle T_g^2)$ for several times of
  integration. The time is in units of the mean collision time. Note that the
  time when a stationary value of the rescaled cumulant is reached is much
  larger than the characteristic time of the system (the collision time).}
\end{minipage}
\hfill
\begin{minipage}[t]{.46\linewidth}
  \includegraphics[width=1 \textwidth,clip=true]{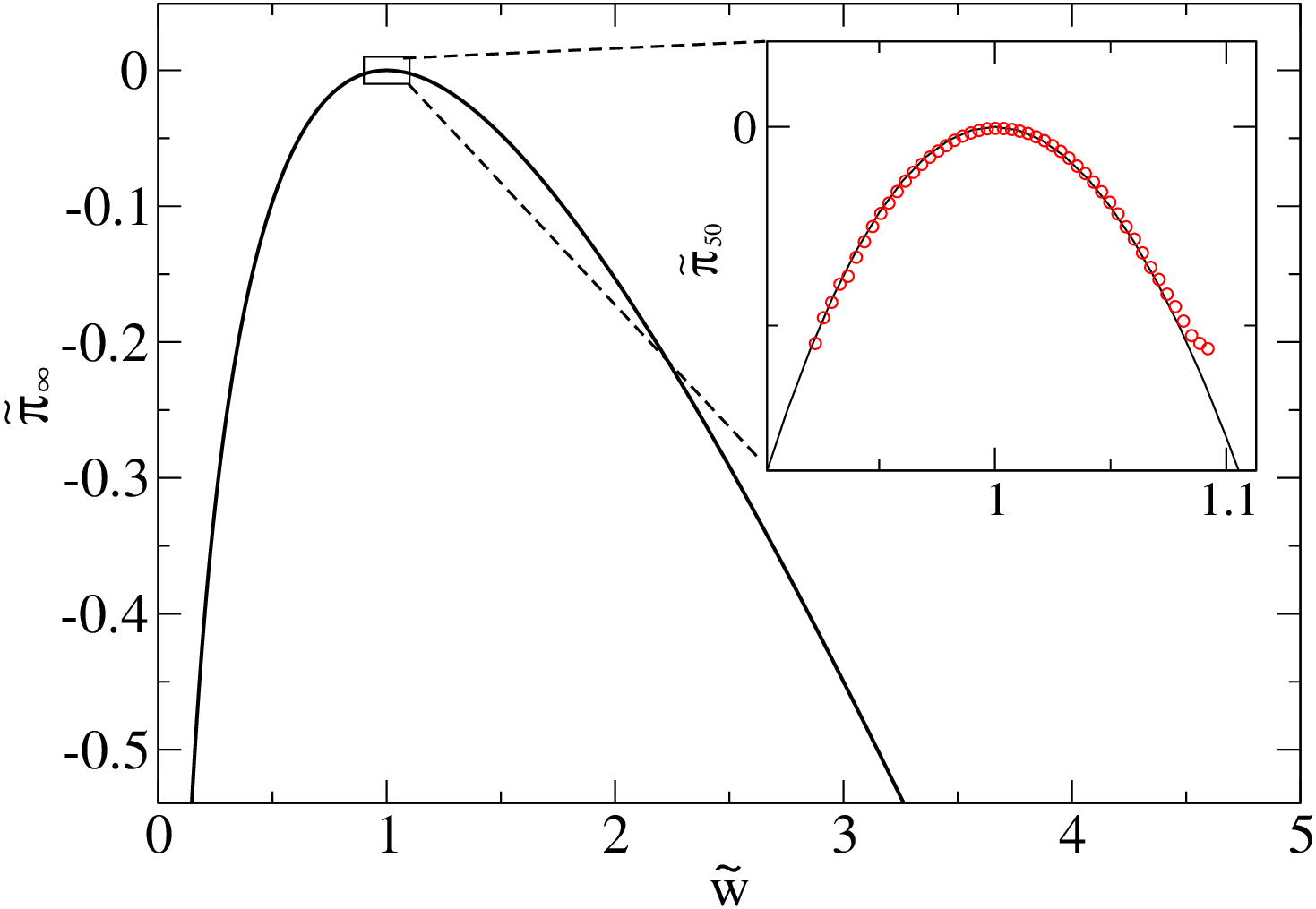}
\caption{\label{piorfwnumeric} Numerical measurement of $\pit_t$ for a time of
  50 collisions per particle (when a stationary value for the rescaled
  cumulant is reached).}
\end{minipage}
\end{figure}

The main conclusion is that no appreciable departure from the $\lambda^2$
truncation is observed at this level of resolution. Much larger statistics are
required to probe the very high energy tails of $p(w,t)$. Further numerical
insights make evident that the small times used to check the GC Relation ($t$
smaller or equal than $3$ mean free times) are far from the time where the
asymptotic large deviation scaling starts working. In
figure~\ref{thirdcumulant} we show indeed the numerical measure of the third
cumulant of ${\cal W}(t)$ rescaled by the first cumulant, varying the
integration time $t$. The time of saturation is of the order of $\sim 50$ mean
free times. The saturation value is in very good agreement with the value
predicted by our theory, eq.~\eqref{4cumulants}. Note that this value is not
at all trivial, since the third cumulant for a Gaussian distribution is zero.
At that time the measurable $\pi_t(w)$ is shown in figure~\ref{piorfwnumeric},
rescaled by $\langle w \rangle$. The accessible range of values from a
numerical simulation is dramatically poor and we think it is already
remarkable to have obtained a good measure of the third cumulant with such a
resolution.

The reason for a verification at small times of the GC formula is the
following: near $w=0$ the pdf of $w$ is almost a Gaussian. In the Gaussian
case we immediately get $\pi_t(w)-\pi_t(-w)=\beta_{eff}w$ with $\beta_{eff}=2
\langle {\cal W}(t) \rangle/\langle {\cal W}(t)^2 \rangle_c$. The first two
cumulants at small times are easily obtained considering an uncorrelated
sequence of energy injection, obtaining $\langle {\cal W}(t) \rangle/t
=N\Gamma d$ and $\langle {\cal W}(t)^2 \rangle_c/t=\langle (\sum_i
\mathbf{F}_i^{th} \cdot \mathbf{v}_i)^2 \rangle_c =2N\Gamma d T_g$. Then the
value $\beta_{eff}=1/T_g$ is unavoidable. In this case the GCFR observed is
nothing else than the Green-Kubo (or Einstein) relation, which is known to be
valid for driven granular gases: $\langle {\cal W}(t)^2 \rangle/t=2T_g\langle
{\cal W}(t) \rangle/t$~\cite{puglisi02,fdtvulpio}.  Small deviations from a Gaussian 
appear, in
first approximation, as small deviations from the slope $1/T_g$, but the
straight line behavior is robust since the first non-linear term of
$\pi_t(w)-\pi_t(-w)$ is not $w^2$ but $w^3$~\cite{aumaitre01}.

Numerical simulations of the Inelastic Maxwell Model have been performed with a
Direct Simulation Monte Carlo analogous to the one used in the Hard Spheres
model. The Maxwell gas is a kinetic model due to Maxwell, who observed that a
pair potential proportional to $r^{-2(d-1)}$, $r$ being the distance between
two interacting particles, gives rise to a great simplification of the
collision integral \cite{maxwell67}. In fact this kind of interaction makes
the collision frequency velocity independent. It must be noted that when the
inelasticity of the particles is considered, this model looses its straight
physical interpretation, but it nevertheless keeps its own interest. The
collision integral is analytically simpler than the hard particles model and
preserves the essential physical ingredients in order to have qualitatively
the same phenomenology. In the recent development of granular gases this
kinetic model has been extensively investigated
\cite{baldassarri02,ben-naim02c,ben-naim03,ernst02,bobylev00}. Thanks to the
simplifications present in this model, we are able to improve the number of
collected data by more than a factor of ten.  The distributions of the
injected power $p(w,t)$ are shown in figure~\ref{fig:max:pdf} for some choices
of the restitution coefficient $\alpha$. The driving amplitude $\Gamma$ has
been changed in order to keep constant the stationary granular temperature
$T_g$.  In figure~\ref{fig:max:pdf_ratio} we have displayed the deviations
from the Gaussian of $P({\cal W},t)$. The non-Gaussianity of $P({\cal W},t)$
is highly pronounced, but again it is striking only in the positive branch of
the pdf.  We have tried, with success, a fit with a fourth order polynomial,
which is consistent with the usual truncation of the Sonine expansion to the
second Sonine polynomial.

Finally, in figure~\ref{fig:max:gc}, we have attempted a check of the
Gallavotti-Cohen fluctuation relation. The relation seems to be systematically
violated. This appears in two points: 1) the right-left ratio of the large
deviation function is not a straight line; 2) the best fitting line has a
slope which is larger than $1/T_g$. The ``curvature''(and the deviation from
the $1/T_g$ line) increases with decreasing values of $\alpha$, indicating
that the inelasticity is the cause of the deviation from the Gallavotti-Cohen
relation. It should be noted that to achieve this result we have collected
more than $4 \times 10^{10}$ independent values of ${\cal W}(t)$, so that the
statistics of the negative large deviations could be clearly displayed.

\begin{figure}
\begin{minipage}[t]{.46\linewidth}
  \includegraphics[width=1 \textwidth,clip=true]{pdf_differentalpha.eps}
\caption{$p(w,t)\equiv t P(wt,t)$ for different values of $\alpha$ (at fixed
  constant temperature $T_g$) in the Driven Inelastic
  Maxwell Model measured at a time $t$ equal to $1$ mean free time. The dashed
  lines are Gaussian distributions with the same mean and same variance. These
  distributions have been obtained with $\sim 4 \times 10^{10}$ independent
  values of ${\cal W}(t)$.}\label{fig:max:pdf}
\end{minipage}
\hfill
\begin{minipage}[t]{.46\linewidth}
  \includegraphics[width=1 \textwidth,clip=true]{pdfratio_differentalpha.eps}
\caption{$p(w,t)$ (at $t$ equal to $1$ mean free time) divided by a Gaussian
  with same average and same variance for different values of $\alpha$ (at
  fixed constant temperature $T_g$) in the Driven Inelastic Maxwell Model. The
  light dashed lines represent a fit with a polynomial of fourth order.}\label{fig:max:pdf_ratio}
\end{minipage}
\end{figure}

\begin{figure}
\begin{center}
  \includegraphics[height=6cm,clip=true]{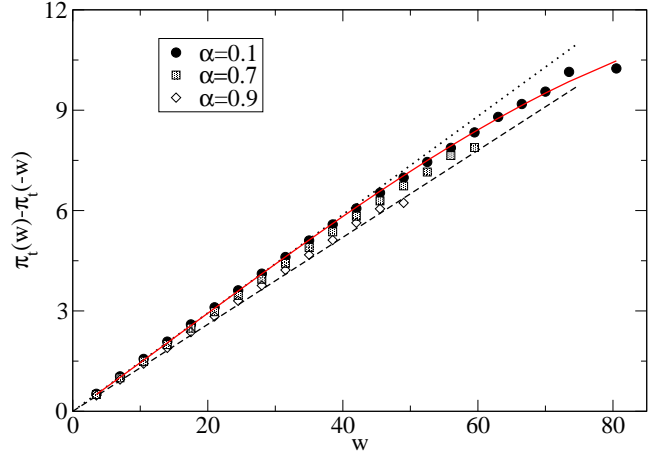}\\
\end{center}
\caption{({\bf Color online}). Finite time check of Gallavotti-Cohen 
  relation for the injected power (with $t$ equal to $1$ mean free time), i.e.
  $\pi_t(w)-\pi_t(-w)$ vs. $w$, in a numerical simulation of the Driven
  Inelastic Maxwell Model with $N=50$, and different values of $\alpha$ (the
  driving amplitude $\Gamma$ has been rescaled in order to fix the granular
  temperature $T_g$).  The dashed curve is a straight line with slope
  $\beta=1/T_g$. The dotted curve is a straight line obtained fitting the
  $\alpha=0.1$ data points until $w=45$, useful as a guide for the eye. The
  thin (red) solid curve is a fit with a cubic ($0.28w + 5.6\cdot 10^{-4} w^2
  - 1.1\cdot 10^{-5} w^3$).} \label{fig:max:gc}
\end{figure}

\subsection{The boundary driven gas of inelastic hard disks}

In a recent experiment on vibrated granular gases~\cite{feitosa04} it has
been argued that the statistics of the power injected on a subsystem by the
rest of the gas fulfills the Fluctuation Relation (FR) by Gallavotti and
Cohen. The experiment was performed by putting in a two-dimensional vertical
box $N$ disks of glass and submitting the container to a strong vertical
vibration. We have reproduced the experiment by means of a Molecular Dynamics
(MD) simulations of inelastic hard disks, observing perfect agreement with the
experimental results and obtaining a deeper insight into the system. The
main difference of this model with respect to the previous ``homogeneously driven''
model is that the external energy source is located at the two horizontal (top
and bottom) walls. This boundary driving mechanism leads to the development of
spatial inhomogeneities and the appearance of internal currents.

\begin{figure}[htbp]
\begin{flushleft}
\makebox[5.8cm][l]{\includegraphics[height=5.5cm]{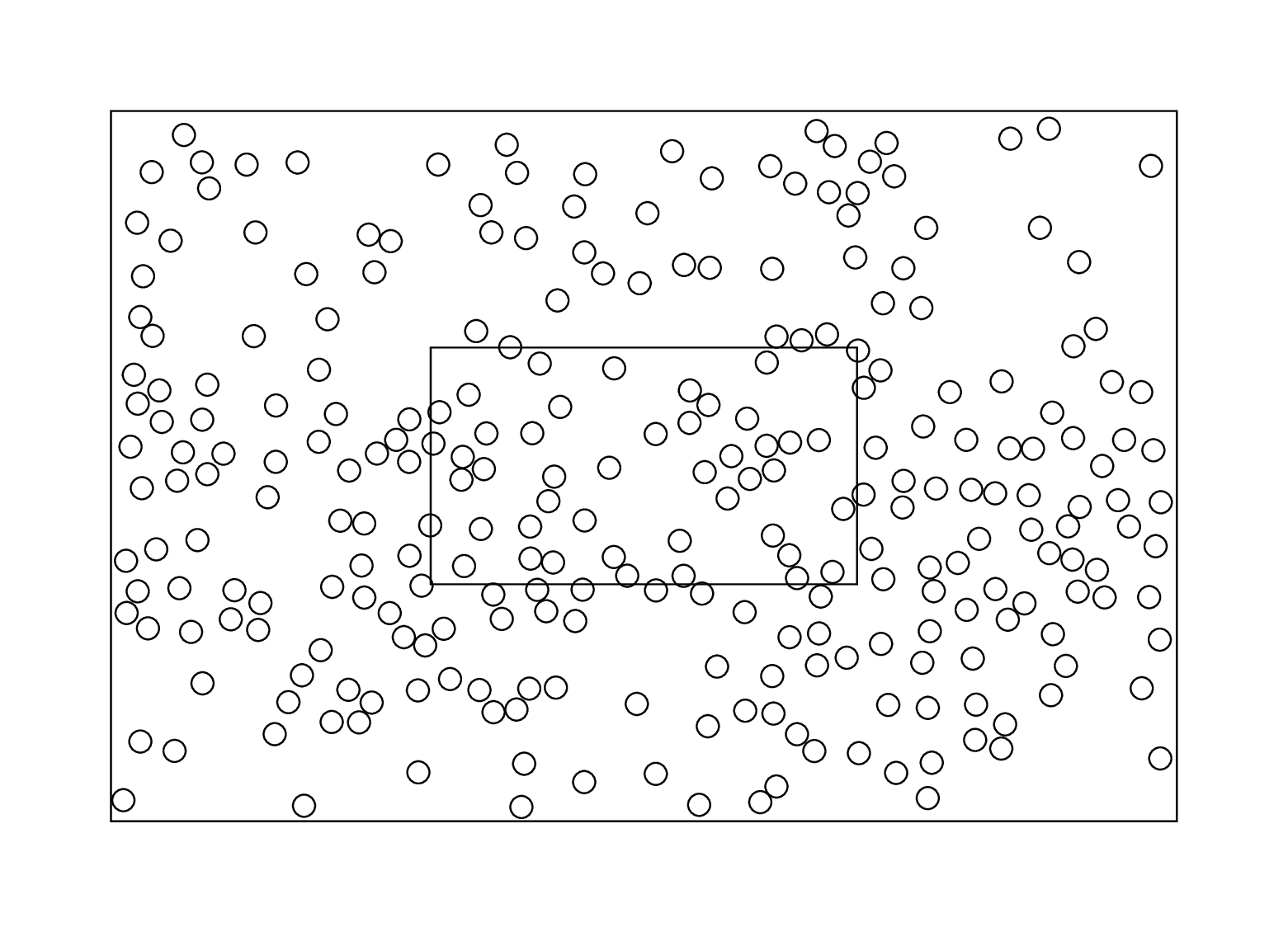}}
\makebox[5.5cm][r]{\includegraphics[height=5.5cm]{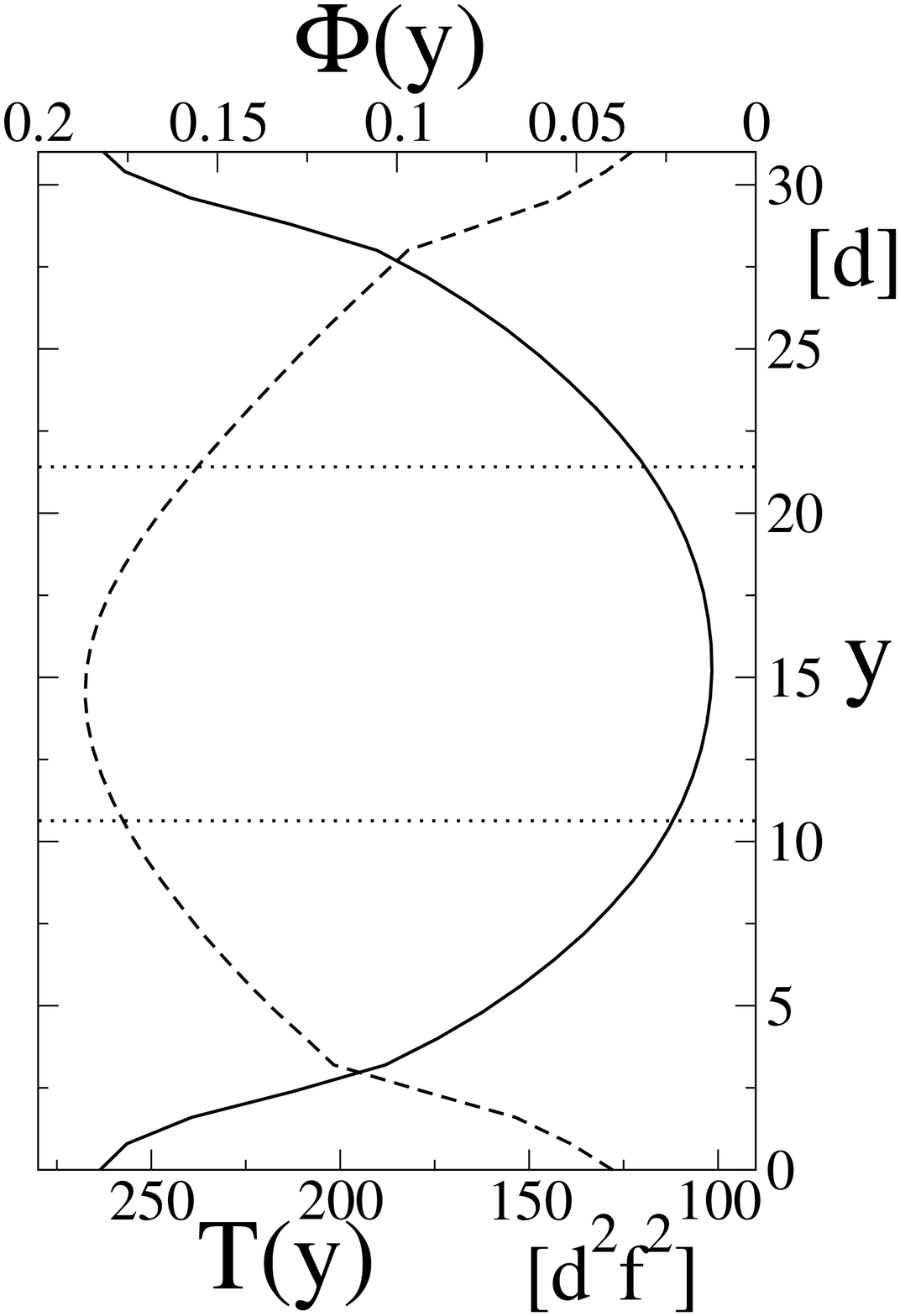}}
\end{flushleft}
\caption{Left: Snapshot of the system considered for MD simulations,
  with the inner region marked by the solid rectangle. Right:
  Corresponding vertical profiles of density ($\Phi(y)$, dashed line)
  and temperature ($T(y)$, solid line). The dotted lines mark the
  bottom and top boundaries of the inner region. Here $N=270$ and
  $\alpha=0.9$. The mean free path is $\sim 5.7d$.
 \label{fig:MD}}
\end{figure}

The event driven MD simulations have been performed for a system of
$N$ inelastic hard disks with restitution coefficient $\alpha$,
diameter $\sigma$ and mass $m=1$.  The vertical $2D$ box of width $L_x=48\sigma$
and height $L_y=32\sigma$ is shaken by a sinusoidal vibration with
frequency $f$ (period $\tau_{box}=1/f$) and amplitude $2.6\sigma$. 
Collisions with
the elastic walls inject energy and allow the system to reach a
stationary state.  We have checked that possible inelastic collisions
with the walls hardly affect the results. Gravity -- set to
$g=-1.7\sigma f^2$ in order to be consistent with the experiment-- has a
negligible influence on the measured quantities.  We have varied the
restitution coefficient from $0.8$ up to $0.99$ (glass beads yield on
average $\alpha \approx 0.9$) and the total area coverage from $0.138$
(i.e.  $N=270$) up to $0.32$ ($N=620$). In figure~\ref{fig:MD}-left a
snapshot of the system is shown.  During the simulations the main
physical observables are statistically stationary.  The local area
coverage field $\Phi(x,y)$ and the granular temperature field $T(x,y)$
(defined in $2D$ as the local average kinetic energy per particle) are
almost uniform in the horizontal direction, apart from small layers
near the side walls.  In figure~\ref{fig:MD}-right the profiles
$\Phi(y)=(1/L_x)\int dx \Phi(x,y)$ and $T(y)=(1/L_x)\int dx T(x,y)$
are shown to be symmetric with respect to the bottom and the top of
the box.  Following the experimental procedure, we have focused our
attention on a ``window'' in the center of the box, fixed in the
laboratory frame, of width $2L_x/5$ and height $L_y/3$, marked in
figure~\ref{fig:MD}-left. Apart from the negligible change of potential
energy due to gravity, the total kinetic energy of the particles
inside the window, changes during a time $\tau$ because of two
contributions: $\Delta K_\tau= Q_\tau - I_\tau$ where $Q_\tau$ is the
kinetic energy transported by particles through the boundary of the
window (summed when going-in and subtracted when going-out) and
$I_\tau$ is the kinetic energy dissipated in inelastic collisions
during time $\tau$.  For several values of $\tau$ we have measured, as
in the experiments, $Q_\tau$ which is related to the kinetic
contribution to the heat flux (we checked that inclusion of the
collisional contribution, even if non small~\cite{herbst04}, does not change
the picture).  With $N=270$ and $\alpha=0.9$ the characteristic times
are the mean free time $\tau_{col} \approx 0.47\tau_{box}$, the
diffusion time across the window $\tau_{diff}=0.82\tau_{box}$ and the
mean time between two subsequent crossings of particles (from outside
to inside) $\tau_{cross}\approx 0.039\tau_{box}$.

\begin{figure}[htbp]
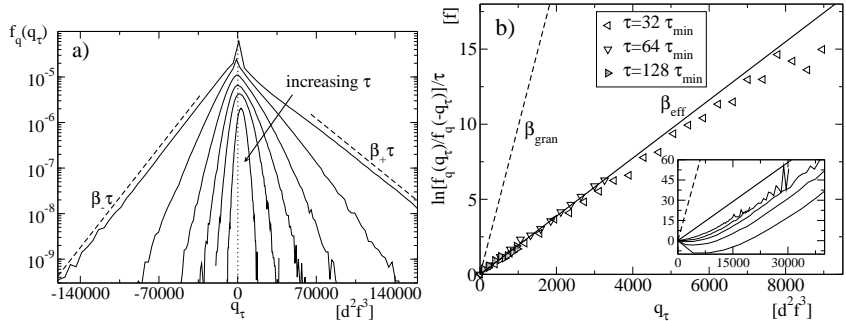

\includegraphics[width=5.5cm,clip=true]{pdfMD.eps}
\includegraphics[width=5.5cm,clip=true]{gc.eps}
\caption{{\bf a)} pdfs of injected power $f_q(q_\tau)$ from MD
  simulations for different values of $\tau=(1,2,4,8,16,32)\times
  \tau_{min}$ with $\tau_{min}=0.015 \tau_{box}$. Here $N=270$ and
  $\alpha=0.9$. The distributions are shifted vertically for
  clarity.  The dashed lines put in evidence the exponential tails of
  the pdf at $\tau=\tau_{min}$.  {\bf b) } plot of
  $(1/\tau)\log[f_q(q_\tau)/f_q(-q_\tau)]$ vs. $q_\tau$ from MD
  simulations (same parameters as above) at large values of $\tau$. The
  solid curve is a linear fit (with slope $\beta_{eff}$) of the data
  at $\tau=128 \tau_{min}$. The dashed line has a slope
  $\beta_{gran}=1/T_{gran}$. In the inset the same graph is shown for
  small values of $\tau=(1,2,4,8) \times \tau_{min}$ (from bottom to
  top).\label{fig:pdfMD}}
\end{figure}

We define the injected power as $q_\tau=Q_\tau/\tau$ and two relevant
probability density functions (pdfs): $f_Q(Q_\tau)$ and $f_q(q_\tau)$.
figure~\ref{fig:pdfMD}a) shows $f_q(q_\tau)$ for different values of $\tau$. A
direct 
comparison with Fig.~3 of Ref.~\cite{feitosa04} suggests a fair agreement
between simulations of inelastic hard disks and the experiment. The pdfs are
strongly non-Gaussian and asymmetric, becoming narrower as $\tau$ is
increased. At small $\tau$ a strong peak in $q_\tau=0$ is visible.  More
interestingly, $f_q(q_\tau)$ at small values of $\tau$ has two different
exponential tails, i.e.  $f_q(q_\tau) \sim \exp(\mp\beta_\pm\tau q_\tau)$ when
$q_\tau \to \pm\infty$ with $\beta_->\beta_+$.  The peak and the exponential
tails at small $\tau$ are observed also in the experiment (see Fig.~3
of~\cite{feitosa04}) and in similar simulations~\cite{aumaitre01}. In
figure~\ref{fig:pdfMD}b) we display $\log[f_q(q_\tau)/f_q(-q_\tau)]/\tau$ vs.
$q_\tau$, which is equivalent to the graph of $\pi_\tau(q_\tau)-\pi_\tau(-q_\tau)$ vs.
$q_\tau$ where $\pi_\tau(q_\tau)=\log[f_Q(\tau q_\tau)]/\tau$.   From
figure~\ref{fig:pdfMD} it appears that at large values of $\tau$,
$\pi_\tau(q_\tau)-\pi_\tau(-q_\tau)$ is linear with a $\tau$-independent slope $\beta_{eff}
\neq 1$. We have measured $\beta_{eff}$ with various choices of the restitution
coefficient $\alpha$ and of the covered area fraction finding similar results.
Ref. \cite{feitosa04} reports $\beta_{eff} T_{gran} \sim 0.25$
where $T_{gran}$ is the mean granular temperature in the observation window.
Similar values are measured in our MD simulations. At area fraction $13.8\%$
and $\alpha=0.9$ we have
$\beta_{eff} T_{gran} \approx 0.23
$. At fixed $\alpha$
and increasing area fraction, $\beta_{eff} T_{gran}$ slightly increases, as in the
experiment.
As $\alpha \to 1$ the slope $\beta_{eff}$ decreases. At $\alpha=1$ (without gravity
and external driving) the distribution of $Q_\tau$ is symmetrical and
$\beta_{eff}=0$, indicating that $1/\beta_{eff}$ is not a physically relevant
temperature concept. Interestingly, it appears that $\beta_{eff}$
is a non hydrodynamic quantity: different systems may show the {\em same}
density and temperature profiles, with very different values of 
$\beta_{eff}$.

We now adopt a coarse-grained description of the experiment which is
able to entirely capture the observed phenomenology. The measured flow of
energy is given by
\begin{equation}
Q_\tau=\frac{1}{2}\left(\sum_{i=1}^{n_+} v_{i+}^2-
\sum_{i=1}^{n_-} v_{i-}^2\right),
\end{equation}
where $n_-$ ($n_+$) is the number of particles leaving (entering) the window
during the time $\tau$, and $v_{i-}^2$ ($v_{i+}^2$) are the squared moduli of
their velocities. In order to analyze the statistics of $Q_{\tau}$ we take
$n_-$ and $n_+$ as Poisson-distributed random variables with average
$\omega\tau$, where $\omega$ corresponds to the inverse of the crossing time
$\tau_{cross}$. In doing so we neglect correlations among particles entering
or leaving successively the central region. A key point, supported by direct
observation in the numerical experiment, lies in the assumption that the
velocities ${\bf v}_{i+}$ and ${\bf v}_{i-}$ come from populations with
different temperatures $T_+$ and $T_-$ respectively. Indeed, compared with the
population entering the central region, those particles that leave it have
suffered on average more inelastic collisions, so that $T_-<T_+$.  Finally we
assume Gaussian velocity pdfs.  
Within such a framework,
the distribution $f_Q(Q_\tau)$ of $Q_\tau$ can be studied analytically. Here
it is enough to recall that $\frac{1}{2}\sum_{i=1}^n v_{i}^2$, in $D$
dimensions, if each component of ${\bf v}_i$ is independently
Gaussian-distributed with zero mean and variance $T$, is a stochastic variable
$x$ with a distribution $\chi_{n,T}(x)=f_{1/T,Dn/2}(x)$, where
$f_{\alpha,\nu}(x)$ is the Gamma distribution, and whose generating function
reads $\tilde{\chi}_{n,T}(z)=\left(1-Tz\right)^{-Dn/2}$~\cite{feller}.  It is
then straightforward to obtain the generating function of $Q_\tau$ in the form
$\tilde{f}_Q(z)=\exp [\tau\mu(z)]$ with
\begin{equation}
\mu(z)=\omega\left(-2+(1-T_+z)^{-D/2}+(1+T_-z)^{-D/2}\right).
\end{equation}
We observe that $\tilde{f}_Q(z)$ has two poles in $z=\pm1/T_\pm$ and
two branch cuts on the real axis for $z>1/T_+$ and $z<-1/T_-$. From $\mu(z)$
we immediately obtain the cumulants of $f_Q(Q_\tau)$ through the formula
$\langle Q^n \rangle_c=\tau\frac{d^n}{dz^n}\mu(0)$.

For $\tau \to \infty$ the large deviation theory states that $f_Q(Q_\tau) \sim
\exp(\tau \pi_\infty(Q_\tau/\tau))$ and $\pi_\infty(q)$ can be obtained from $\mu(z)$ through
a Legendre transform, i.e. $\pi_\infty(q)=\smash{\underset{z}{max}}(\mu(z)-qz)$. 
The study of
the  singularities of $\mu(z)$ reveals the behavior of the large deviation
function $\pi_\infty(q)$ for $q \to \pm\infty$. It can be seen that
\begin{align} \label{tails}
\pi_\infty(q) \sim -\frac{q}{T_+} \;\;(q \to \infty), \;\;\;
\pi_\infty(q) \sim \frac{q}{T_-} \;\;(q \to -\infty).
\end{align}
We emphasize however that it is almost impossible to appreciate these tails in
simulations and in experiments, since the statistics for large values
of $q$ and $\tau$ is very poor.

A Gallavotti-Cohen-type relation~\cite{gallavotti95,lebowitz99,kurchan98},
e.g. $\pi_\infty(q)-\pi_\infty(-q)=\beta q$ for any $q$ and an arbitrary value of
$\beta$ would imply $\mu(z)=\mu(\beta -z)$. One can see
that such a $\beta$ does not exist, i.e. the fluctuations of $Q_\tau$
do not satisfy a Gallavotti-Cohen-like relation.  The observed
linearity of the graph
$\log[f_q(q_\tau)/f_q(-q_\tau)]/\tau=\pi(q_\tau)-\pi(-q_\tau)$
vs. $q_\tau$ can be explained by the same observation pointed out in the
previous subsection: at large values of $\tau$ it is extremely
difficult, in simulations as well as in experiments, to reach large
values of $q$, while for small $q$, $\pi(q)-\pi(-q) \approx
2\pi'(0)q+o(q^3)$, i.e. a straight line with a slope
$\beta_{eff}=2\pi'(0)$ is likely to be observed. It has been already
shown~\cite{farago02} that in dissipative systems deviations from the FR
can be hidden by insufficient statistics at high values of $q$.  The
knowledge of $\mu(z)$ is useful to predict this slope.  
At large $\tau$, $\pi'(0)\approx\Pi'(0)=-z^*(0)$ where $z^*(q)$ is the
value of $z$ for which $\mu(z)-qz$ is extremal. This gives
\begin{equation}
\beta_{eff}=\frac{T_+^{\delta}-T_-^{\delta}}{T_+^{\delta+1}+T_-^{\delta+1}}
\quad \hbox{with} \quad
\delta=\frac{2}{2+D}.
\label{eq:beta}
\end{equation}
When $T_+=T_-$
(i.e. if $\alpha=1$)
$\beta_{eff}=0$.  We emphasize
that $\beta_{eff}$ does not depend upon $\omega$.  We have compared
with success these predictions with the numerical and experimental results,
measuring the temperatures $T_+$ and $T_-$ in the simulation. 

What happens for small values of $\tau$? We note that $\tilde{f}_Q(z)$ has the
form $\exp(\tau\mu(z))$ for {\em any} value of $\tau$ and not only for large
$\tau$. Therefore at small $\tau$ one can expand the exponential, obtaining
$\tilde{f}_Q(z) \sim 1+\omega\tau\left(-2+(1-T_+ z)^{-D/2}+\right .$ $\left
  .(1+T_- z)^{-D/2}\right)$.  This immediately leads to an analytical
expression for $f_Q(Q_\tau)=\text{const} \times
\delta(Q_\tau)+\chi_{1,T_+}(Q_\tau)+\chi_{1,T_-}(-Q_\tau)$, which fairly
accounts for the strong peak which is observed in the experiment and in the
simulations, and predicts exponential tails for $f_Q(Q_\tau)$: $\chi_{1,T}(x)
\propto x^{D/2-1}\exp(-x/T)$ so that $\beta_+=1/T_+$ and $\beta_-=1/T_-$.
This suggests an experimental test of this theoretical approach: the measure
at small values of $\tau$ of the slopes of the exponential tails of
$f_Q(Q_\tau)$ should coincide with a direct measure of $T_+$ and $T_-$.
However, we point out that the values of $\beta_+$ and $\beta_-$ obtained by
fitting the tails in the hard disks simulation, using values as small as
$\tau=0.00015 \tau_{box}$ yield estimates of $T_+$ and $T_-$ which are larger
(by a factor $\sim 1.6$) than those found by a direct measure.  This
disagreement brings the limits of such a simple two-temperature picture to the
fore. In the simulation and in the original experiment the measured injected
energy is indeed the sum of several different contributions, namely $Q_\tau
\approx Q_\tau^{xx} + Q_\tau^{xy} + Q_\tau^{yx} + Q_\tau^{yy}$ where
$Q_\tau^{ij}$ is the kinetic energy transported by the $i$ component of the
velocity by particles crossing the boundary through a wall perpendicular to
direction $j$. Two main differences with the simplified interpretation given
above arise: a) there are {\em two} couples of temperatures, i.e.
$T_+^x,T_-^x$ as well as $T_+^y,T_-^y$~\cite{brey98c,mcnamara98b,blair03}; b)
the diagonal contributions $Q_\tau^{jj}$ are sums of squares of velocities
whose distribution is not a Gaussian but is $\sim v \exp(-v^2/T)$, since the
probability of crossing is biased by the velocity itself. The calculation of
$f_Q(Q_\tau)$ is still feasible, with qualitatively similar results.

\section{The dynamics of a tracer particle as a non-equilibrium Markov process}
\label{tracer}

In the search for a quantity that is, more rigorously, related to the
``entropy production'' in a granular gas,  we consider in this section the
projection of the dynamics of the gas onto that of a tracer particle, which is
easier since it is equivalent to a jump Markov process. We are interested in
the dynamics of a tracer granular particle in a homogeneous and dilute gas of
grains which is driven by an unspecified energy source. The requirements are
that the gas is dilute, spatially homogeneous and time 
translational invariant. 
The gas is characterized by its velocity probability density
function $P(\mathbf{v})$ which, for the sake of simplicity, will be
considered of the form $P(\mathbf{v})=\frac{1}{(2\pi T)^{d/2}}\exp{ \left( -
    \frac{v^2}{2T}\right)}(1+a_2S_2^{d}(v^2/2T))$, (where $S_2^d$ is the
$d$-dimensional second Sonine polynomial already defined in~\ref{soninep}).
The gas is therefore parametrized by its temperature $T$ and its second
Sonine coefficient $a_2$ which measures its non-Gaussianity.

The linear Boltzmann equation for the tracer, in generic dimension
$d$, reads:
\begin{multline} \label{boltzmann}
\frac{\dd P_*(\mathbf{v},t)}{\dd t}=\frac{1}{\ell}\int \dd \mathbf{v}_1 \int \dd \mathbf{v}_2 \int' \dd \bo |(\mathbf{v}_1-\mathbf{v}_2) \cdot \bo |
P_*(\mathbf{v}_1) P(\mathbf{v}_2) \times \\ \times \left\{\delta\left(\mathbf{v}-\mathbf{v}_1+\frac{1+\alpha}{2}[(\mathbf{v}_1-\mathbf{v}_2)\cdot \bo] \bo\right)
-\delta(\mathbf{v}-\mathbf{v}_1)\right\}
\end{multline}
where $P_*(v)$ is the velocity pdf of the test particle and the primed
integral again indicates that the integration is performed on all angles that satisfy
$(\mathbf{v}_1-\mathbf{v}_2) \cdot \bo >0$. The mean free path $\ell$ appears
in front of the collision integrals. In the following (when not stated
differently) we will put $\ell=1$, which can be always obtained by a rescaling
of time.

We rewrite the above equation~\eqref{boltzmann} as a Master
equation for a Markov jump process~\cite{puglisi05b}:

\begin{equation} \label{eq:markov}
\frac{\dd P_*(\mathbf{v},t)}{\dd t}=\int \dd \mathbf{v}_1 P_*(\mathbf{v}_1) K(\mathbf{v}_1, \mathbf{v}) - \int \dd \mathbf{v}_1 P_*(\mathbf{v}) K(\mathbf{v}, \mathbf{v}_1).
\end{equation}
The transition rate $K(\mathbf{v}, \mathbf{v}')$ of jumping from
$\mathbf{v}$ to $\mathbf{v}'$ is given by the following formula:
\begin{equation}
\label{eq:K}
K(\bv,\bv')=\left(\frac{2}{1+\alpha}\right)^2
\frac{1}{\ell} |\Delta \bv|^{2-d} 
\int \dd \bv_{2 \tau} P[\mathbf{v}_2(\mathbf{v},\mathbf{v}',\bv_{2 \tau})],
\end{equation}
where $\Delta \bv=\bv' - \bv$ denotes the change of velocity of the
test particle after a collision. The vectorial function $\bv_2$ is defined as
\begin{equation}
\mathbf{v}_2(\mathbf{v},\mathbf{v}',\bv_{2\tau})=v_{2\sigma}
(\mathbf{v},\mathbf{v}')\bs(\mathbf{v},\mathbf{v}')+
 \bv_{2\tau}  \label{eq:Kb},
\end{equation}
where $\bs(\mathbf{v},\mathbf{v}')$ is the unitary vector parallel to $\Delta
\bv$, while $\bv_{2 \tau}$ is entirely contained in the $(d-1)$-dimensional
space perpendicular to $\Delta \bv$ (i.e.  $\bv_{2 \tau} \cdot \Delta \bv
=0$).  This implies that the integral in expression~\eqref{eq:K} is
$(d-1)$-dimensional. Finally, to fully determine the transition rate
(\ref{eq:K}), the expression of $v_{2 \sigma}$ is needed:
\begin{equation}
v_{2 \sigma}(\bv , \bv') = \frac{2}{1+ \alpha} |\Delta \bv| 
+ \bv \cdot \bs \,\,\,.
\end{equation}

\subsection{Detailed balance}

Here, we obtain a simple expression for the ratio between
$K(\mathbf{v},\mathbf{v}')$ and $K(\mathbf{v}',\mathbf{v})$. When
exchanging $\mathbf{v}$ with $\mathbf{v}'$ the unitary vector $\bs$ changes sign.
Furthermore one has that $v_{2 \sigma}(\bv, \bv') \ne v_{2 \sigma}(\bv', \bv)$.
From all these considerations and from equation~\eqref{eq:K} one obtains
immediately:
\begin{equation}
\frac{K(\mathbf{v},\mathbf{v}')}{K(\mathbf{v}',\mathbf{v})}=
\frac{\int \dd \bv_{2\tau}P[\mathbf{v}_2(\bv, \bv')]}{\int \dd \bv_{2\tau}P[\mathbf{v}_2(\bv', \bv)]}
\equiv  \frac{P[v_{2 \sigma} (\bv, \bv')]}{P[v_{2 \sigma} (\bv', \bv)]}\,\,\,.
\end{equation}
We note that this ratio depends only on the choice of the pdf of the gas, $P$, and not
on the other parameters (such as $\alpha$). However in realistic
situations (experiments or Molecular Dynamics simulations) $P$ is not a free
parameter but is determined by the choice of the setup (e.g. external
driving, material details, geometry of the container, etc.).

Introducing the short-hand notation $v_{2 \sigma}=v_{2 \sigma}(\bv, \bv')$, $v_{2 \sigma}'=v_{2 \sigma}(\bv', \bv)$ and $v_{\sigma}^{( \prime )}=\bv^{( \prime )} \cdot \bs$,
we also note that

\begin{equation}
(v_{2\sigma}')^2=v_{2\sigma}^2+(v_{\sigma}+v_{\sigma}')^2-2v_{2\sigma}(v_{\sigma}+v_{\sigma}')\,\,,
\end{equation}
from which it follows that 

\begin{equation} \label{eq:delta2}
\Delta_2 =(v_{2\sigma})^2-(v_{2\sigma}')^2=-\Delta-2\frac{1-\alpha}{1+\alpha}\Delta=-\frac{3-\alpha}{1+\alpha}\Delta\,\,,
\end{equation}
where $\Delta=v_{\sigma}^2-(v_{\sigma}')^2 \equiv
|v|^2-|v'|^2$, i.e. the kinetic energy lost by the test-particle
during one collision. When $\alpha=1$ then $\Delta_2=
-\Delta$ (energy conservation).
From the above considerations it follows that 

\begin{itemize}

\item
in the {\bf Gaussian} case, it is found

\begin{equation}
\log\frac{K(\mathbf{v},\mathbf{v}')}{K(\mathbf{v}',\mathbf{v})}=\frac{\Delta}{2T} +2 \frac{1-\alpha}{1+\alpha}\frac{\Delta}{2T}=\frac{3-\alpha}{1+\alpha}\frac{\Delta}{2T}
\end{equation}

\item
in the {\bf First Sonine Correction} case, it is found

\begin{equation} \label{ratio_sonine}
\log\frac{K(\mathbf{v},\mathbf{v}')}{K(\mathbf{v}',\mathbf{v})}=\frac{3-\alpha}{1+\alpha}\frac{\Delta}{2T}+\log\frac{\left\{1+a_2S_2^{d=1}\left[\frac{\left(\frac{2}{1+\alpha}(v_{\sigma}'-v_{\sigma})+v_{\sigma}\right)^2}{2T}\right]\right\}}{\left\{1+a_2S_2^{d=1}\left[\frac{\left(\frac{2}{1+\alpha}(v_{\sigma}-v_{\sigma}')+v_{\sigma}'\right)^2}{2T}\right]\right\}}
\end{equation}

\end{itemize}

In the case where $P(v)$ is a Gaussian with temperature $T$, it is immediate
to observe that
\begin{equation}
P_*(\mathbf{v})K(\mathbf{v},\mathbf{v}')=
P_*(\mathbf{v}')K(\mathbf{v}',\mathbf{v})
\end{equation}
if $P_*$ is equal to a Gaussian with temperature
$T'=\frac{\alpha+1}{3-\alpha}T \le T$. This means that there is a Gaussian
stationary solution of equation~\eqref{eq:markov} (in the Gaussian-bulk case),
which satisfies detailed balance. The fact that such a Gaussian with a
different temperature $T'$ is an exact stationary solution was known
from~\cite{martin99}. It thus turns out that detailed balance is
satisfied, even out of thermal equilibrium. Of course this is an artifact of
such a model: it is highly unrealistic that a granular gas yields a Gaussian
velocity pdf. As soon as the gas velocity pdf $P(v)$ ceases to be Gaussian,
detailed balance is violated, i.e. the stationary process performed by the
tracer particle is no more in equilibrium within the thermostatting gas.  We
will see in section~\ref{noneq} how to characterize this departure from
equilibrium.

\subsection{Action functionals} \label{noneq}

From the previous section we have learnt that the dynamics of the velocity of
a tracer particle immersed in a granular gas is equivalent to a Markov process
with well defined transition rates. This means that the velocity of the tracer
particle stays in a state $\mathbf{v}$ for a random time $t\geq 0$ distributed
with the law $r(\mathbf{v}) e^{-r(\mathbf{v})t} \dd t$ and then makes a
transition to a new value $\mathbf{v}'$ with a probability
$r(\mathbf{v})^{-1}K(\mathbf{v},\mathbf{v}')$, with $r(\mathbf{v})=\int \dd
\mathbf{v}' K(\mathbf{v},\mathbf{v}')$. At this point it is interesting to ask
about some characterization of the non-equilibrium dynamics, i.e. of the
violation of detailed balance, which we know to happen whenever the
surrounding granular gas has a non-Gaussian distribution of velocity.

To this extent, we define two different action functionals,
following~\cite{lebowitz99}:

\begin{subequations} \label{entropy}
\begin{align}
  W(t)&= \sum_{i=1}^{n(t)} \log \frac{K(\mathbf{v}_i \to
    \mathbf{v}_i')}{K(\mathbf{v}_i' \to \mathbf{v}_i)}\\
  \overline{W}(t)&= \log\frac{P_*(\mathbf{v}_1)}{P_*(\mathbf{v}_{n(t)}')}+
  \sum_{i=1}^{n(t)} \log \frac{K(\mathbf{v}_i \to
    \mathbf{v}_i')}{K(\mathbf{v}_i' \to
    \mathbf{v}_i)} \\ &\equiv\log\frac{\mathcal{P}(\mathbf{v}_1 \to \mathbf{v}_2
    \to ... \to \mathbf{v}_{n(t)})}{\mathcal{P}(\mathbf{v}_{n(t)} \to
    \mathbf{v}_{n(t)-1} \to ... \to \mathbf{v}_1)} \label{entropyb}
\end{align}
\end{subequations}
where $i$ is the index of collision suffered by the tagged particle,
$\mathbf{v}_i$ is the velocity of the particle before the $i$-th collision,
$\mathbf{v}_i'$ is its post-collisional velocity, $n(t)$ is the total number
of collisions in the trajectory from time $0$ up to time $t$, and $K$ is the
transition rate of the jump due to the collision. Finally, we have used the
notation $\mathcal{P}(\mathbf{v}_1 \to \mathbf{v}_2 \to ... \to \mathbf{v}_n)$
to identify the probability of observing the trajectory $\mathbf{v}_1 \to
\mathbf{v}_2 \to ... \to \mathbf{v}_n$. The quantities $W(t)$ and
$\overline{W}(t)$ are different for each different trajectory (i.e. sequence
of jumps) of the tagged particle. Note that the first term
$\log\frac{P_*(\mathbf{v}_1)}{P_*(\mathbf{v}_{n(t)}')}$ in the definition of
$\overline{W}(t)$, eq.~\eqref{entropyb}, is non-extensive in time.  The two above
functionals have the following properties:

\begin{itemize}
  
\item $W(t) \equiv 0$ if there is exact symmetry, i.e. if $K(\mathbf{v}_i \to
  \mathbf{v}_{i+1})=K(\mathbf{v}_{i+1} \to \mathbf{v}_i)$ (e.g. in the
  microcanonical ensemble); $\overline{W}(t) \equiv 0$ if there is detailed
  balance (e.g. any equilibrium ensemble);
  
\item we expect that, for large enough $t$, for almost all the trajectories
  $\lim_{s \to \infty} W(s)/s = \lim_{s \to \infty} \overline{W}(s)/s=\langle
  W(t)/t \rangle=\langle \overline{W}(t)/t \rangle$; here (since the system
  under investigation is ergodic and stationary) the meaning of $\langle
  \rangle$ is intuitively an average over many independent segments of a
  single very long trajectory;
 
\item for large enough $t$: 1) at equilibrium $\langle W(t) \rangle=\langle
  \overline{W}(t)\rangle=0$; 2) out of equilibrium (i.e. if detailed balance
  is not satisfied) those two averages are positive; we use those
  equivalent averages, at large $t$, to characterize the distance from
  equilibrium of the stationary system;

\item if $S(t)=-\int \dd v P_*(v,t)\log P_*(v,t)$ is the entropy associated
  to the pdf of the velocity of the tagged particle $P_*(v,t)$ at time $t$ 
  (e.g. $-H$ where $H$ is the Boltzmann-H function), then

\begin{equation}
\frac{\dd}{\dd t}S(t)=R(t)-A(t)
\end{equation}

where $R(t)$ is always non-negative, $A(t)$ is
linear with respect to $P_*$ and, finally, $\langle W(t) \rangle
\equiv \int_0^t \dd t' A(t')$. This leads to consider $W(t)$ equivalent to the
contribution of a single trajectory to the total entropy flux. In a stationary
state $A(t)=R(t)$ and therefore the flux is equivalent to the production; this
property has been recognized in~\cite{lebowitz99}.

\item {\bf $FR_W$} (Lebowitz-Spohn-Gallavotti-Cohen fluctuation relation):
  $\pi(w)-\pi(-w)=w$ where $\pi(w)=\lim_{t\to \infty}\frac{1}{t}\log{f_W^t(t
    w)}$ and $f_W^t(x)$ is the probability density function of finding
  $W(t)=x$ at time $t$; at equilibrium the $FR_W$ has no content; note that in
  principle $\pi'(w,t)=\frac{1}{t}\log{f_W^t(t w)}\neq\pi(w)$ at any finite
  time; a generic derivation of this property has been obtained
  in~\cite{lebowitz99}, while a rigorous proof with more restrictive hypothesis
  is in~\cite{maes99}; the discussion for the case of a Langevin equation is
  in~\cite{kurchan98}.
  
\item {\bf $FR_{\overline{W}}$} (Evans-Searles fluctuation relation):
  $\overline{\pi}(w,t)-\overline{\pi}(-w,t)=w$ where
  $\overline{\pi}(w,t)=\frac{1}{t}\log{f_{\overline{W}}^t(t w)}$ and
  $f_{\overline{W}}^t(x)$ is the probability density function of finding
  $\overline{W}(t)=x$ at time $t$; at equilibrium the $FR_{\overline{W}}$ has
  no content; this relation is derived in~\cite{lebowitz99}; the analogy
  between this relation and the Evans-Searles fluctuation
  relation~\cite{evans94,evans02} has been put forward in~\cite{puglisi05b}

\end{itemize}

A detailed numerical study~\cite{puglisi05b} of the fluctuations of $W(t)$ and
$\overline{W}(t)$ in this model has shown on the one hand that, out of equilibrium
(i.e. when the surrounding gas is non-Gaussian), the $FR_{\overline{W}}$ is
always satisfied. On the other hand the $FR_W$ is always violated, even if it
was expected on the basis of the arguments given in~\cite{lebowitz99}. The
difference between the two functionals defined in~\eqref{entropy} is a term
which is non-extensive in time, but which has fluctuations whose distribution
has exponential tails and therefore, in principle, can contribute to the large
deviation function of $W(t)$~\cite{bordi,visco2t}. Such a failure of a large time Fluctuation
Relation, which is much more pronounced in the near-to-equilibrium cases, is
similar to that observed in other
systems~\cite{evans05,farago02,vanzon03,bonetto05}

\section{Conclusions}
\label{conclusions}

The study of the fluctuations of global physical quantities in a granular gas
is at its very beginning. In the lack of a general rigorous theory in the
framework of non equilibrium statistical mechanics, experiments and numerical
simulations are the main source of results, together with few exact analytical
calculations. In this review of recent
results~\cite{visco05,puglisi05,visco05b,puglisi05b,viscoepj} we have indicated some
routes that have been followed, focusing on two global quantities (total
energy and energy injection rate) that are of interest in nowadays physics of
non-equilibrium
systems~\cite{bramwell98,brey05,aumaitre01,aumaitre04,farago02,farago04}.  On
one hand we have shown that total energy fluctuations have a pdf that strongly
depends on the model considered.  We have also pointed out that definitive
inferences about the presence of correlations, starting from the observation
of ``anomalous'' pdfs of total energy, must be drawn with caution, since the
lack of spatial or temporal translational invariance may play a major role. On
the other hand we have presented a method to calculate the large deviation
function of injected power in a granular gas: this method strongly suggests
the disappearance of a negative branch in such large deviation function. This
result is a direct consequence of the time-irreversibility of inelastic
collisions: injected power fluctuations are dominated at large times by the
energy dissipated in collisions, which is always positive. Finally we have
sketched a recipe to obtain a quantity related to time-reversal asymmetry
(i.e. violation of detailed balance) whose large deviations can be both
positive and negative. This quantity has the advantage of being measurable in
experiments, but the disadvantage of not having an obvious ``macroscopic''
counterpart. It contains in fact information on the non-Gaussianity of the
velocity pdf of the gas. This consideration is in our opinion the main open
issue in this study.

\bibliographystyle{alpha}
\bibliography{granulates}


\printindex
\end{document}